\documentclass[11pt,journal]{IEEEtran}
\pdfoutput=1
\usepackage[bookmarks=false]{hyperref}
\usepackage{subfigure}
\usepackage{tabularx}
\usepackage{multirow}
\usepackage{color}
\usepackage[pdftex]{graphicx}
\usepackage{lineno}
\usepackage{xspace}
\usepackage{amsmath}
\usepackage{amssymb}
\usepackage{ifthen}

\usepackage[ansinew]{inputenc}
\usepackage[english]{babel}

\newcommand{\ACKIN} {\ensuremath{{\cal S}_{in }}\xspace}
\newcommand{\ACKOUT}{\ensuremath{{\cal S}_{out}}\xspace}

\newcommand{\TERM}[2]{\ensuremath{t({#1},{#2})}}

\newcommand{\gf}[1][]%
{\ensuremath{\ifthenelse{\equal{#1}{}}{\mathbb{F}_2}{\mathbb{F}_2^{#1}}}\xspace}

\newcommand{\INV}{\ensuremath{(0, 0)\xspace}}

\begin{document}
\title{A Secure Asynchronous FPGA Architecture, Experimental Results and Some Debug Feedback}
            
\author{Sumanta~Chaudhuri, Sylvain~Guilley, Philippe~Hoogvorst and Jean-Luc~Danger\\
Institut T\'EL\'ECOM / T\'EL\'ECOM Paris, CNRS -- LTCI (UMR 5141)\\
46 rue Barrault, 75\,634 PARIS Cedex 13, FRANCE. \\
Taha~Beyrouthy, Alin~Razafindraibe, Laurent~Fesquet and Marc~Renaudin\\
\textsc{TIMA} Laboratory (INPG), CIS group\\
46 avenue F\'elix Viallet, 38\,031 GRENOBLE, FRANCE.}

\maketitle
            
\begin{abstract} 
This article presents an asynchronous FPGA architecture for implementing cryptographic algorithms secured against 
physical cryptanalysis. We discuss the suitability of asynchronous reconfigurable architectures for such applications before
proceeding to model the side channel and defining our objectives.
The logic block architecture is presented in detail. We discuss
several solutions for the interconnect architecture, and how these solutions can be ported to other flavours of interconnect (i.e. single driver).
Next We discuss in detail a high speed asynchronous configuration chain architecture used to configure our asynchronous FPGA with simulation results, and we present a $3 \times 3$ prototype 
FPGA fabricated in 65~nm CMOS.
Lastly we present experiments to test the high speed asynchronous configuration chain and evaluate how far our objectives have been achieved with proposed solutions,
and we conclude with emphasis on complementary FPGA CAD algorithms, and the effect of CMOS variation on Side-Channel Vulnerability.
\end{abstract}
            
%
%

\section{Introduction}
\label{sec:Introduction}

Cryptography is a mean to defend against potential attackers, notably to protect confidentiality, integrity or secure authentication,
whereas cryptanalysis is about the challenge to retrieve hidden information.
There are no known mathematical cryptanalysis methods which 
can decrypt standard cryptographic algorithms like AES in a reasonable amount of time and space, assuming that the cryptanalyst 
has access to both plain-text and encrypted messages. However all such algorithms are implemented with some physical process, that leak information.
An access to this information makes the job of the cryptanalyst much easier.
These kinds of information leakage from 
physical processes are commonly known as side-channel leakage.

For the purpose of this article, we divide cryptanalysis broadly in two categories: mathematical and physical. 
In this article, we assume that concerned cryptographic algorithms are secure at the mathematical level, and we specifically address
the issue of physical cryptanalysis and countermeasures. Physical cryptanalysis can again be of two types, namely active and passive. Injecting 
faults to perturb the physical implementation is an example of active attacks, whereas attacks based on measuring power consumption / electromagnetic (EM) radiation 
are examples of passive attacks, commonly known as Side-Channel Attacks (SCAs).

Physical cryptanalysis has been demonstrated to be effective against various standard algorithms, and on various platforms in recent times.
Researchers have shown that side-channel attacks can be mounted on standard cryptographic algorithms like DES~\cite{kocher-timing_attacks}, AES~\cite{aes-attack}, RSA~\cite{kocher-timing_attacks}.
References~\cite{oop-ches03,soqp-fpl04} provide with the details of such attacks on FPGA implementations whereas various attacks~\cite{tiri-dac07} has been reported on ASIC implementation.
A widely known SCA is DPA (Differential Power Analysis)~\cite{kocher-dpa_and_related_attacks}, which exists in various forms~\cite{Practical_Template_Attacks} and concerns the information
leaked through supply current peaks. Attacks which exploit the Electromagnetic Emissions (EMA)~\cite{EM} from the hardware, constitute another major branch 
of Side-Channel Attacks. The attacks on RSA, which use the difference in execution time, as their major source of information have also been reported,
and these are commonly known as Timing Attacks~\cite{kocher-timing_attacks}. The reader could as well find a comprehensive report of active attack details in~\cite{fault-attacks}.

Now then, who's at risk? A very evident answer should be banking applications. Credit cards use algorithms similar to RSA for authentication,
and 2-key Triple DES for the challenge~\cite{EMV}. Wholesale frauds on systems which rely on smart cards for their security (e.g. Pay-TV) could well be
a target of such attacks. 
Mounting a side-channel attack calls for considerable expertise and high-resolution equipments.
So such techniques are prone to be used when there is a considerable gain. A major threat could be intellectual property protection because of this reason.
The attacker can easily gain access to piracy protection devices embedded in commercial systems, and steal the IPs. All the more reason to incorporate
side channel resistance into these systems.

In the rest of article we will move in a top-down fashion. 
Section~\ref{sec:Asynchronous Circuits} presents the asynchronous circuits and protocols, and the salient points of this technology. 
Section~\ref{sec:Side Channel Attacks} provides a brief overview and classification of side-channel attacks, the assumed models and 
a classification of countermeasures.
In Section~\ref{sec:Suitability of Asynchronous FPGA for Physical Cryptanalysis Resistance} we list the features of asynchronous reconfigurable circuits that make them especially suitable for security applications.
Once the  reader gains an understanding of where this article is situated among these vast interacting domains, we provide a model for the side channel in
section~\ref{sec:Modelling The Side Channel} and set our objectives. We present the  logic block architecture of our asynchronous FPGA in section~\ref{sec:Logic Block Architecture}. 
Section~\ref{sec:Routing Architecture} addresses the issue of interconnect design, which makes up the most of the area in an FPGA and 
section~\ref{sec:Single Driver Architecture} presents the method to port these solutions to the new single driver architecture. 
In Section~\ref{sec:Prototype} we present a prototype asynchronous FPGA and section~\ref{sec:Experiments} presents the evaluation 
of the proposed solutions based on experiments. Section~\ref{sec:Conclusion} presents the conclusions from this research effort.

\section{Asynchronous Circuits}
\label{sec:Asynchronous Circuits}
\begin{figure}
\centering
{
\subfigure[Synchronous]{\includegraphics[width=0.5\textwidth]{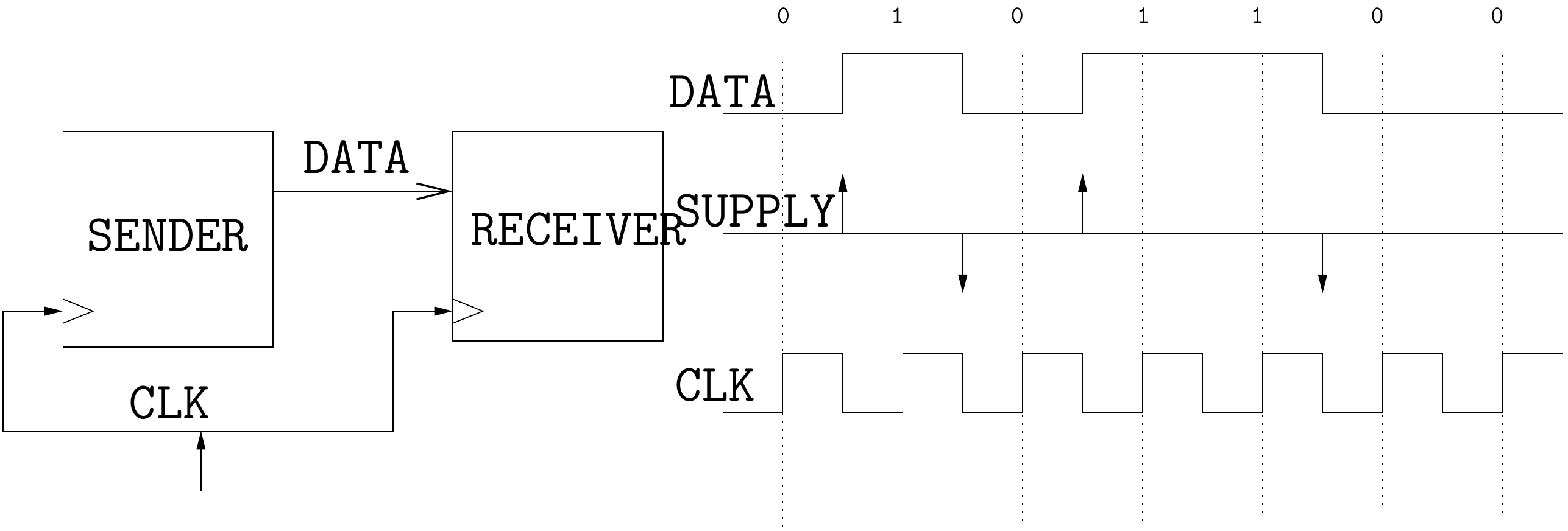}
\label{fig:sync}}
\subfigure[Basic Handshake Protocol]{\includegraphics[width=0.5\textwidth]{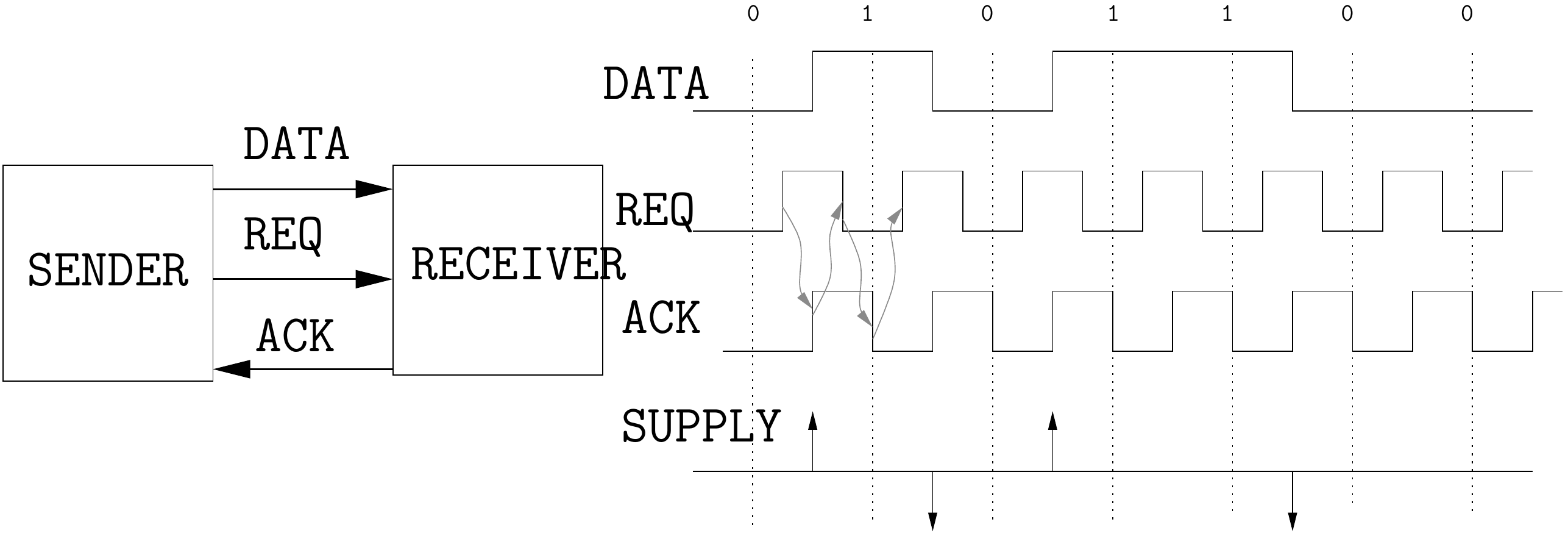}
\label{fig:async1}}
\subfigure[1-out-of-2 4-phase]{\includegraphics[width=0.5\textwidth]{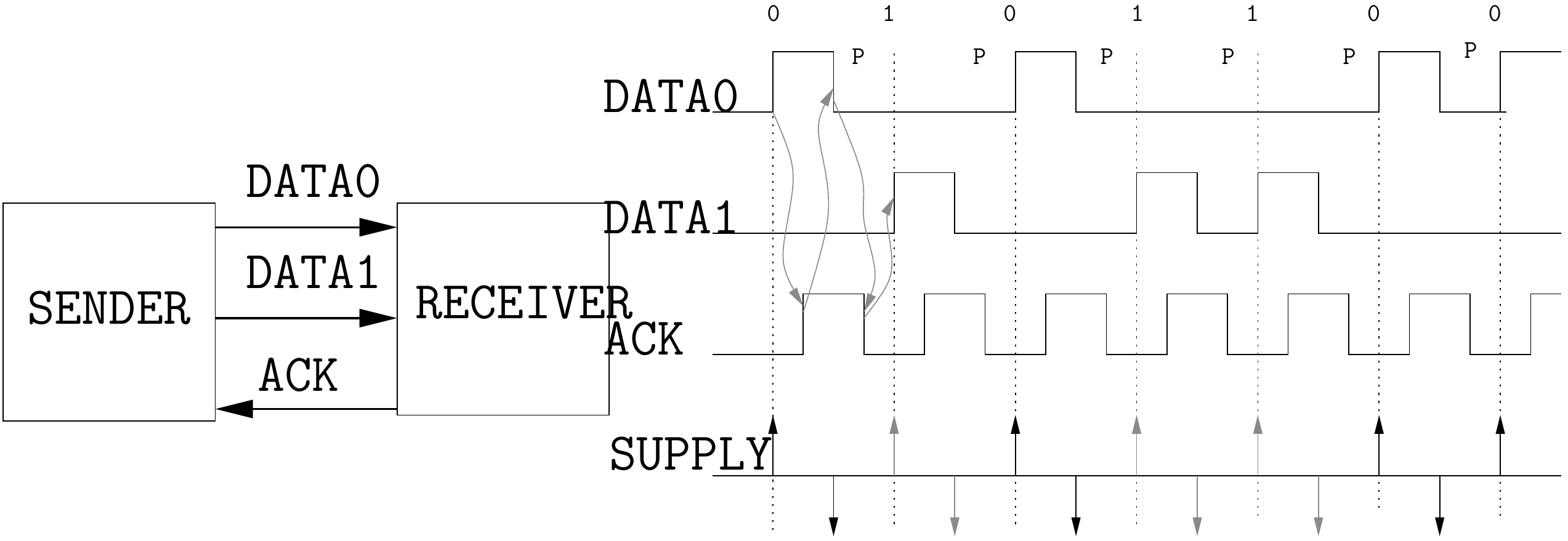}
\label{fig:async2}}
}
\caption{Asynchronous protocols: Figure~\ref{fig:sync} shows the synchronous methodology where data is valid only at the positive edge of the clock signal. Figure~\ref{fig:async1} shows
the handshaking in asynchronous protocol, the arrows show the causality of the request and acknowledgement events.}
\label{fig:async}
\end{figure}
In this section we discuss the key ideas of asynchronous logic. The author welcomes the reader to take a look at the following publications
for more detailed discussions~\cite{Martin90,cortadella97petrify,sutherland}. Figure~\ref{fig:async1} shows the basic asynchronous handshake protocol. The sender puts valid data on the data 
line and sends a request on the REQ line to show the validity of data. Once the receiver has read the data, it asserts the ACK line so that sender 
can put some new data. This basic protocol is formally defined as production rules or Seitz's weak conditions~\cite{Seitz80}.

However this basic scheme is not hazard free. For example, if the line has more delay than the REQ line, it is possible that the receiver gets
the request before the DATA is valid. To avoid such hazards the DATA and REQ are often encoded into a single channel. Most common encodings are $1$-out-of-$n$
encodings similar to one hot codes for finite state machines.
Figure~\ref{fig:async2} depicts the $1$-out-of-$2$ encoding.
Table~\ref{tab:4phase} describes the encoding scheme.
Apart from encodings, the signalling protocol
can also be of various flavours. The popular four phase protocol uses one phase for computation, and one phase to precharge all signals to zero state.
At the opposite,
2-phase protocols (NRZ) does not use precharge.
Details about 2-phase protocols can be found in~\cite{linder96phased,DBLP:journals/cee/MoradiSS09}.

\begin{table}
\begin{center}
\caption{$1$-out-of-$2$, $4$-phase protocol.}
\label{tab:4phase}
\begin{tabular}{|l|p{1cm}|p{1cm}|}
\hline
\texttt{Value}     & \texttt{DATA0} & \texttt{DATA1} \\ \hline
\texttt{'0'}       & \texttt{'1'} & \texttt{'0'} \\
\texttt{'1'}       & \texttt{'0'} & \texttt{'1'} \\
\texttt{Precharge} & \texttt{'0'} & \texttt{'0'} \\
\texttt{Forbidden} & \texttt{'1'} & \texttt{'1'} \\ \hline
\end{tabular}
\end{center}
\end{table}
Since in a asynchronous signalling scheme each event has significance (as opposed to synchronous logic where a glitch can occur without disturbing the 
functionality), the asynchronous protocols are completely glitch free. Glitches are results of unbalanced input path arrival times (unbalanced joins).
In asynchronous circuits, this hazard is taken care of with C-Elements~\cite{shams-cmos_cmuller} at the gate level, however at the transistor level, there is an additional constraint
of forks balanced in delays. This constraint is commonly known as ``isochronous fork'' constraint~\cite{Martin90}, and such asynchronous circuits are called Quasi-Delay Insensitive
(QDI) asynchronous circuits. In this article we assume QDI asynchronous circuits implicitly whenever we discuss asynchronous protocols.

Various asynchronous FPGA architectures proposed in literature often use the properties of asynchronous logic for high performance (high speed, low power, robustness).
Typical architectures divide into several categories: fine grain~\cite{cornell,Tei04,Tei03}, coarse grain~\cite{Mak98,Pay97,Hau92} and GALS~\cite{Gao96}.
Indeed, with asynchronous logic 
glitch-free operation and absence of clock network can substantially reduce power consumption, and slack elasticity~\cite{cornell} of asynchronous can augment the throughput.
The architecture presented in this article has its focus on resistance against physical cryptanalysis, while enjoying the above-mentioned benefits
of being asynchronous.

In the most recently proposed asynchronous FPGA architecture~\cite{cornell,Tei04,Tei03} the logic block is designed assuming the 1-out-2 4-phase protocol, and uses pipelines in the routing switches to increase throughput. Routing segments are a 3-wire bundle (DATA0, DATA1, ACK).
Our logic block architecture is much more fine grain to accommodate a plethora of encoding schemes and styles and the routing architecture is single wires, on which dual-rails are routed together. This is done, keeping in mind the
prototyping role of an FPGA, and flexibility required for dynamic countermeasures to operate. Since we don't know of any future-proof solution to resist 
physical cryptanalysis, this architecture will provide the designer a soft fine grain fabric on which he/she can implement a mix of dynamic and static 
countermeasures pertinent to the application.
In section~\ref{sec:Experiments}, we will discuss the additional cost to be paid for this added flexibility.

\section{Side Channel Attacks}
\label{sec:Side Channel Attacks}
%
Side-Channel Attacks are very similar to Spectroscopy(NMR)
used over the years. While in spectroscopy, patterns in
the light spectrum are used to detect the presence of atoms
and its environments in an unknown substance, in a Side-
Channel Attack the cryptanalyst looks for patterns in the
power consumption or EM emission to detect the unknown key value.

We classify Side-Channel Attacks in two ways. They are either based on the acquisition method:
\begin{itemize}
\item
Supply current measurement (DPA)
\item
EM emission measurement (EMA)
\item
Timing difference measurement (Timing Attack)
\end{itemize}
or based on processing methods:
\begin{itemize}
\item
Correlation based. (The measured traces are correlated with the predicted trace from assumed model)
\item
Template. (The model itself is created from experimental measurements on one sample, which is then used to predict the traces for clone circuits~\cite{Practical_Template_Attacks})
\end{itemize}
We will give a very basic example of how a side-channel attack is carried out.
A broad overview can be found in~\cite{tiri-dac07}.
Referring to Fig.~\ref{fig:sync} we can see that in an unprotected synchronous logic, each change of state of a signal can be distinguished
by a current spike, and no change of state by an absence of current spike in the power supply line. Given this basic behaviour of CMOS circuits a side-channel 
cryptanalyst could proceed in the following fashion:
\begin{itemize}
\item
He finds a signal which is a function of say $N$ bits of the key, and the input message of $M$ bits.
\item
He performs the encryption $2^M$ for each different message value and acquires the power trace of each of them.
\item
He makes $2^N$ key guesses and for each key guess he make current spike predictions for $2^M$ traces.
\item
Among these $2^N$ current spike predictions for $2^M$ encryptions, the one which correlates best with the measured power trace, 
is the correct key guess.
Indeed, the asymptotic prediction matches the real observation only for the correct key guess.
\end{itemize}

\begin{figure}
   \centering
   {\subfigure[Raw power traces after acquisition.]{
   \includegraphics[width=0.4\textwidth]{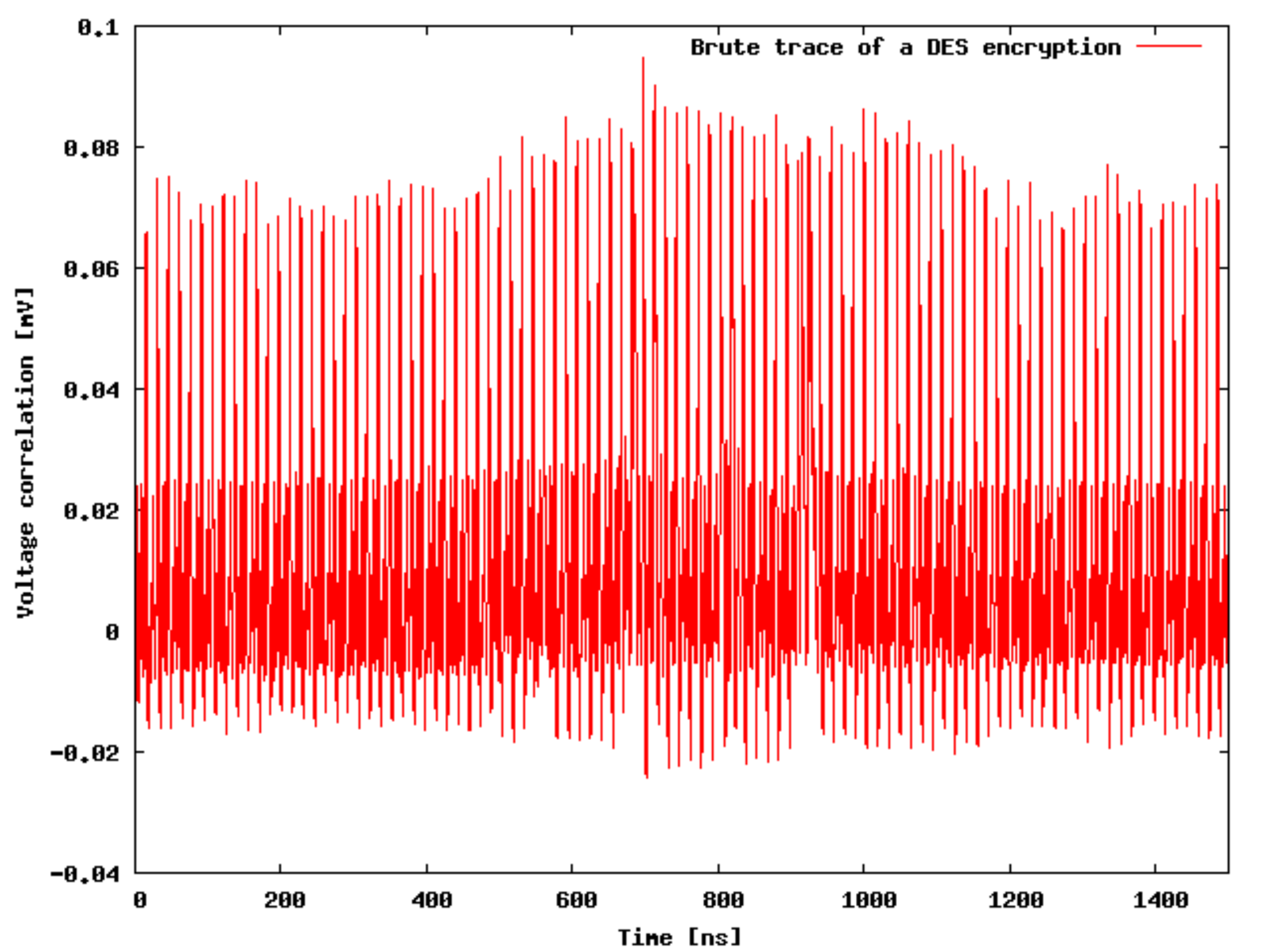}
   \label{fig:scademo1}}
   \subfigure[After Processing.]{
   \includegraphics[width=0.4\textwidth]{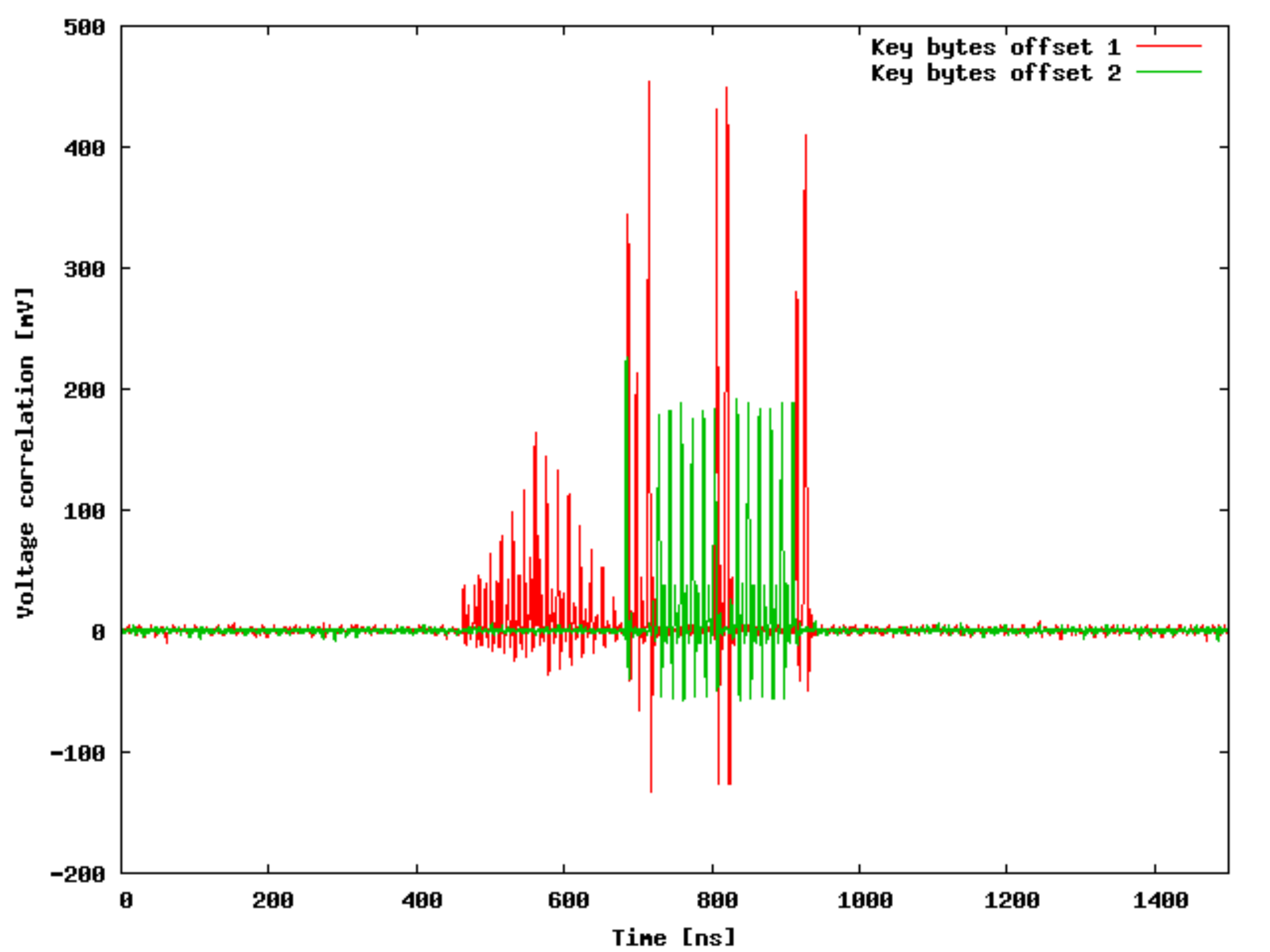}
   \label{fig:scademo2}}}
   \caption{Correlation peaks appear for the right guess: Figure~\ref{fig:scademo1} shows the power consumption traces of a complete DES encryption process. This raw power trace is 
	analysed for 64 possible sub-key guess. Figure~\ref{fig:scademo2} is the resulting waveform for the right guess for two power predictions.}
\end{figure}

The activity of other nodes are not correlated since they are not a function of the targeted message and key bits. These activities of
other nodes appear as noise after the processing of acquired traces. The resistance to physical cryptanalysis, relies on maximizing 
this signal to noise ratio (SNR) either by static or dynamic countermeasures.
Figure~\ref{fig:scademo1} shows the raw power traces after acquisition on an ASIC implementation of DES, and figure~\ref{fig:scademo2} 
shows the appearance of predicted peak when correlated with the right key guess.

\subsection{Countermeasures}
\label{subsec:Countermeasures}
The reported countermeasures to side-channel attacks can be classified as dynamic countermeasures (incorporated at run-time) and static counter 
measures (incorporated at design time).
\subsubsection{Dynamic Countermeasures}
\label{subsubsec:Dynamic}
The principle of dynamic countermeasures is to decorrelate computing from the power supply current, by randomising transitions, commonly known
as ``masking''.
This can be by either precharging signals with random values, or introducing random delays in computing paths~\cite{oswald-fse05,akkar-ches01,SRQ06}. These masking 
techniques are introduced at the algorithmic level.
Details about implementing and attacking such countermeasures can be found in~\cite{DBLP:conf/ches/PeetersSDQ05,schaumont-ches07}.
\subsubsection{Static Countermeasures}
\label{subsubsec:Static}
Static countermeasures rely on producing a constant power consumption profile independent of the data being computed. This is commonly done with 
differential signalling with a precharge to `0', where power consumption profile of one rail hides that of the other. Examples of such countermeasures 
are WDDL~\cite{tiri-date04}, Backend Duplication~\cite{guilley-date04}, or STTL~\cite{raza-patmos07}.
$1$-out-of-$n$ asynchronous signalling also falls in this category.
To mitigate the remaining unbalance of the various rails,
an unpredictable random switching of them can be enforced.
MDPL~\cite{popp-ches05} is a typical example of such a strategy.

\section{Suitability of Asynchronous FPGA for Physical Cryptanalysis Resistance}
\label{sec:Suitability of Asynchronous FPGA for Physical Cryptanalysis Resistance}
In this section we point out to the reader the motivations behind the architecture we are going to discuss. The suitability of asynchronous
circuits for cryptanalysis resistance has already been investigated by~\cite{anderson-async02}.
\begin{itemize}
\item
\textbf{Resistance to Fault Attacks.}
Random introduction of faults stalls the asynchronous circuit~\cite{DBLP:journals/mam/MooreAMTF03,DBLP:journals/tc/MonnetRL06}.
So the cryptanalyst does not receive the encrypted messages with fault syndromes.
To do so, faults have to be injected very carefully, at a precise time and location, which makes the attack considerably difficult.
\item
\textbf{Absence of a Time Reference.}
The absence of a reference signal (i.e. CLK) in an asynchronous circuit, prevents the attacker to assume a precise model for transitions he is
trying to predict, whereas in a synchronous circuit the targeted transitions must occur within the clock cycle. Moreover, the power consumption
of the clock signal is clearly visible in the power trace, and provides an overall idea of the circuit operation.
\item
\textbf{Power Constant Signalling.}
As shown in figure~\ref{fig:sync} the supply current spikes, clearly denote the change of state of the signal, or the absence of a peak denotes
that no changes occurred in the signal value.
On the contrary, for asynchronous $1$-out-of-$2$ signalling (see Fig.~\ref{fig:async2}) each valid signal
value is accompanied by one spike in the supply current. Note that both signals are precharged to a neutral value (``00'') in between the valid data.
This power constant signalling falls well into the category of static countermeasures previously discussed in section~\ref{subsubsec:Static}.
\item
\textbf{Absence of Glitches.}
In synchronous implementations, glitches can occur without disturbing the functionality. Glitches magnify the current spikes shown in figure~\ref{fig:sync}.
Reference~\cite{Fischer05} discusses the effect of glitches on Side-Channel Attacks. As discussed in section~\ref{sec:Asynchronous Circuits} asynchronous circuits can not work in the presence of glitches and 
are consequently less vulnerable.
\item
\textbf{Reconfigurability.}
The motivation to opt for a reconfigurable architecture, rather than a hardwired circuit is firstly to achieve a mix between the dynamic and static countermeasures,
(see section~\ref{subsec:Countermeasures} and \cite{chaudhuri-reconfig06,DBLP:conf/ches/MentensGV08}) depending on the application. Secondly, as a prototyping platform for evaluating various asynchronous styles and/or masking techniques.
\end{itemize}

\section{Modelling the Side Channel}
\label{sec:Modelling The Side Channel}

\subsection{Dynamic Power Consumption Model}
\label{subsec:Dynamic Power consumtion model}
\begin{figure}
\centering
{
\subfigure[Based on positive \& negative step responses.]{\includegraphics[width=0.5\textwidth]{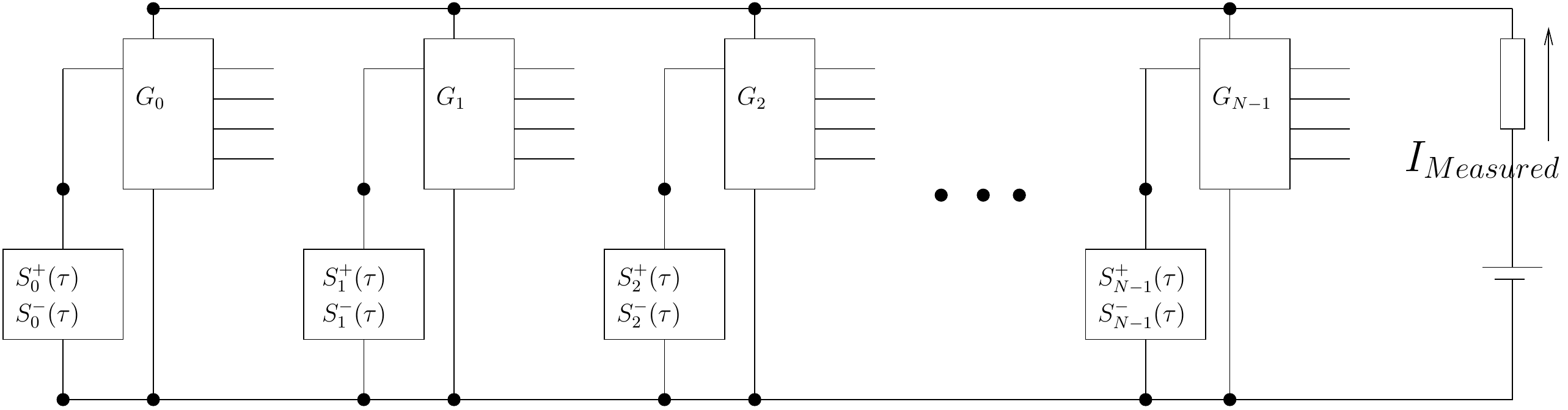}
\label{fig:model1}}
\subfigure[Model for a buffered net.]{\includegraphics[width=0.5\textwidth]{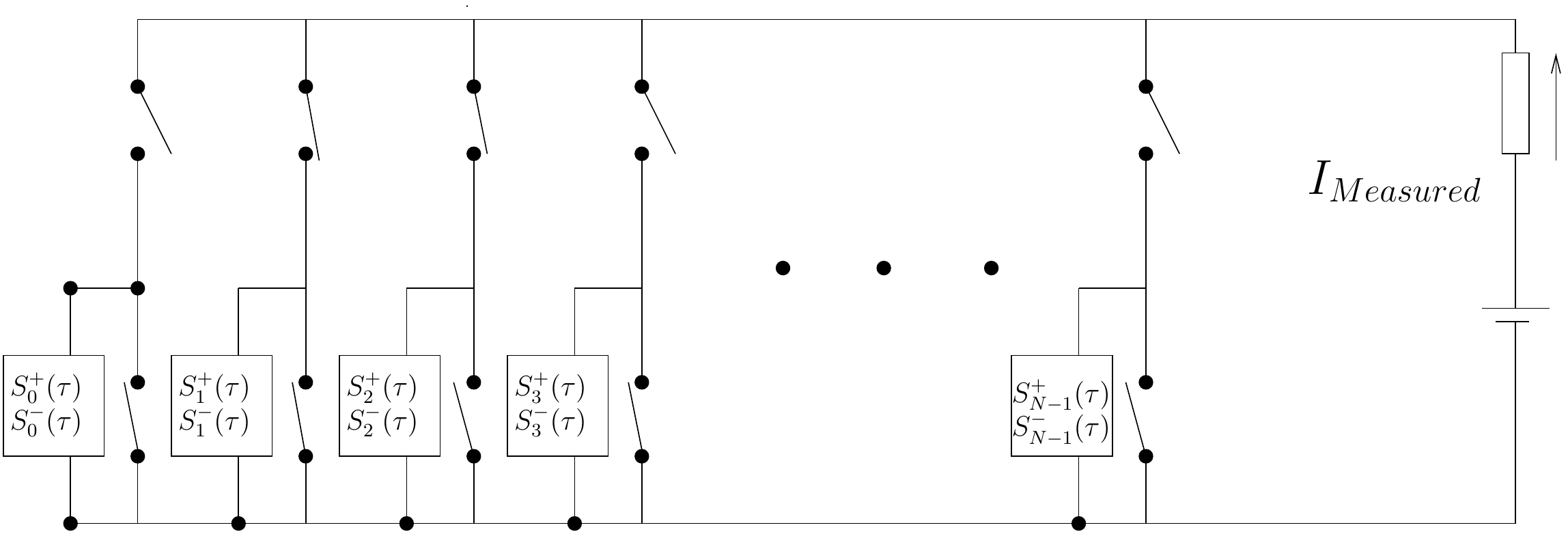}
\label{fig:model}}
}
\caption{Dynamic power consumption model: Figure~\ref{fig:model1} shows the power model of a CMOS logic circuit as the sum of +ve/-ve step responses of the individual gates. Figure~\ref{fig:model} shows
the power consumption model of buffered nets considered as gates with only one input. Unbuffered nets are included in
the capacitance seen by the gate driving that net.}
\label{fig:powermodel}
\end{figure}

At first order, static leakage power in CMOS does not contain any information about the computation being performed and is a constant hence we do not take it into account
for SCA.
Power consumption is proportional to the current charged and discharged from the power supply.

We model dynamic power consumption in CMOS in two levels. 
\begin{itemize}
\item
First the internal power consumption of the gate, which is due to 
charging and discharging of internal nets and transistor short-circuit currents inside the gate. 
\item
Secondly the power consumption of the net driven by the gate
which also includes the input capacitances of the driven gates.
\end{itemize}

Side-Channel Information is in the dynamic current profile of the circuit, thus we need a detailed model for 
this consumption. For this reason, each gate and each net is associated with its step current response (i.e. the contribution of the component
to the current $I_{measured}$ (see fig.~\ref{fig:model1})).

\paragraph{Gate Level}
We consider gates with $N$ inputs. Thus the gate is characterized by its step current response as the input vector undergoes a transition $i_j\rightarrow i_{k}$
while the gate output is open.
\begin{equation}
S^{i_j\rightarrow i_{k}}(\tau)= I_{measured}(\tau)
\end{equation}

and the gate is characterized by the set of all step responses corresponding to each transition.

\begin{displaymath}
\bigcup_{\substack{0<i<2^N-1 \\ 0<j<2^N-1}} S^{i_j\rightarrow i_{k}}(\tau)
\end{displaymath}

\paragraph{Net Level}
Each net has only one input, hence it is characterized by the set which contains its positive and negative step response. 
\begin{displaymath}
\left[S^{0\rightarrow1}(\tau),S^{1\rightarrow0}(\tau)\right]
\end{displaymath}
while the input to the net is a positive or negative step.

We consider both positive and negative step response because the charging and discharging network for the net could be different
in the actual layout.

We model  a buffered net  as depicted in figure~\ref{fig:model}. It is a delayed sum of the step responses of each segment, and the step responses 
for active gates (buffers). This point to point delay is calculated using widely used Elmore Delay model as described in the next section.


\subsubsection{Delay Model}
\label{sec:Delay Model}
To calculate the point to point delay in the above model we use the widely used Elmore delay model~\cite{ASIC}.
The Elmore delay is given by:
\begin{equation*}
\tau_i  =  \sum_{k=1}^{N} C_{k}R_{ik} \,,
\end{equation*}
where $N$ is the number of capacitances in the equivalent network and $R_{ik}$ is given by
\begin{equation*}
R_{ik} = \sum R_j \Rightarrow \left(R_j \in \left[\mathrm{path} \left(S \to i\right) \cap \mathrm{path} \left(S \to k\right) \right] \right)
\end{equation*}

\begin{figure}[t]
\centering
\includegraphics[width=0.45\textwidth]{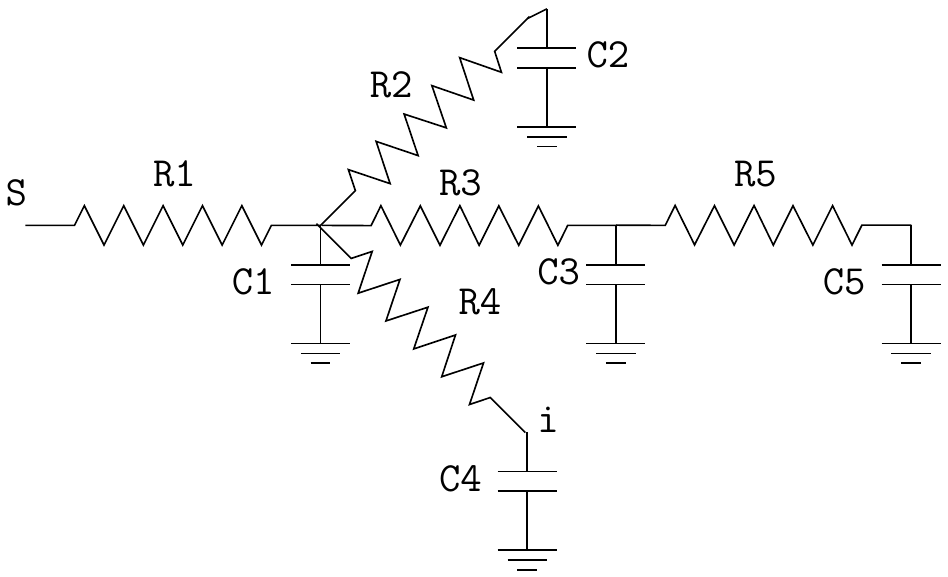}
\caption{Delay model for the FPGA interconnect.}
\label{fig:delay}
\end{figure}

\subsection{Secure Place-Route Objectives}
\label{sec:Secure Place-Route Objectives}
\subsubsection{\textbf{Indiscernability in power consumption}}
\paragraph{Gate Level}

In section~\ref{subsec:Dynamic Power consumtion model} we have modelled the power consumption profile of a gate (a LUT in this case) as a set of step current response for each transition.
Asynchronous logic gates are mapped into two symmetric LUTs as shown in figure~\ref{fig:mapping}. According to the 1-out-of-2 4-phase protocol, only one of these 
two gates will evaluate for each evaluate and precharge cycle. 
To guarantee that the current consumption profile of these two gates are similar, we try to assure that:
\begin{itemize}
\item
for a LUT each path from input to the output is  indiscernible from each other.
\item
each path from the configuration memory point to the output is also indiscernible from each other.
\end{itemize}

If these conditions are met, it is not possible to predict which gate has evaluated, even if the corresponding dual-rails does not use the same
inputs of the corresponding LUTs.

\paragraph{Net Level}
\begin{figure}[t]
\centering
\includegraphics[width=0.45\textwidth]{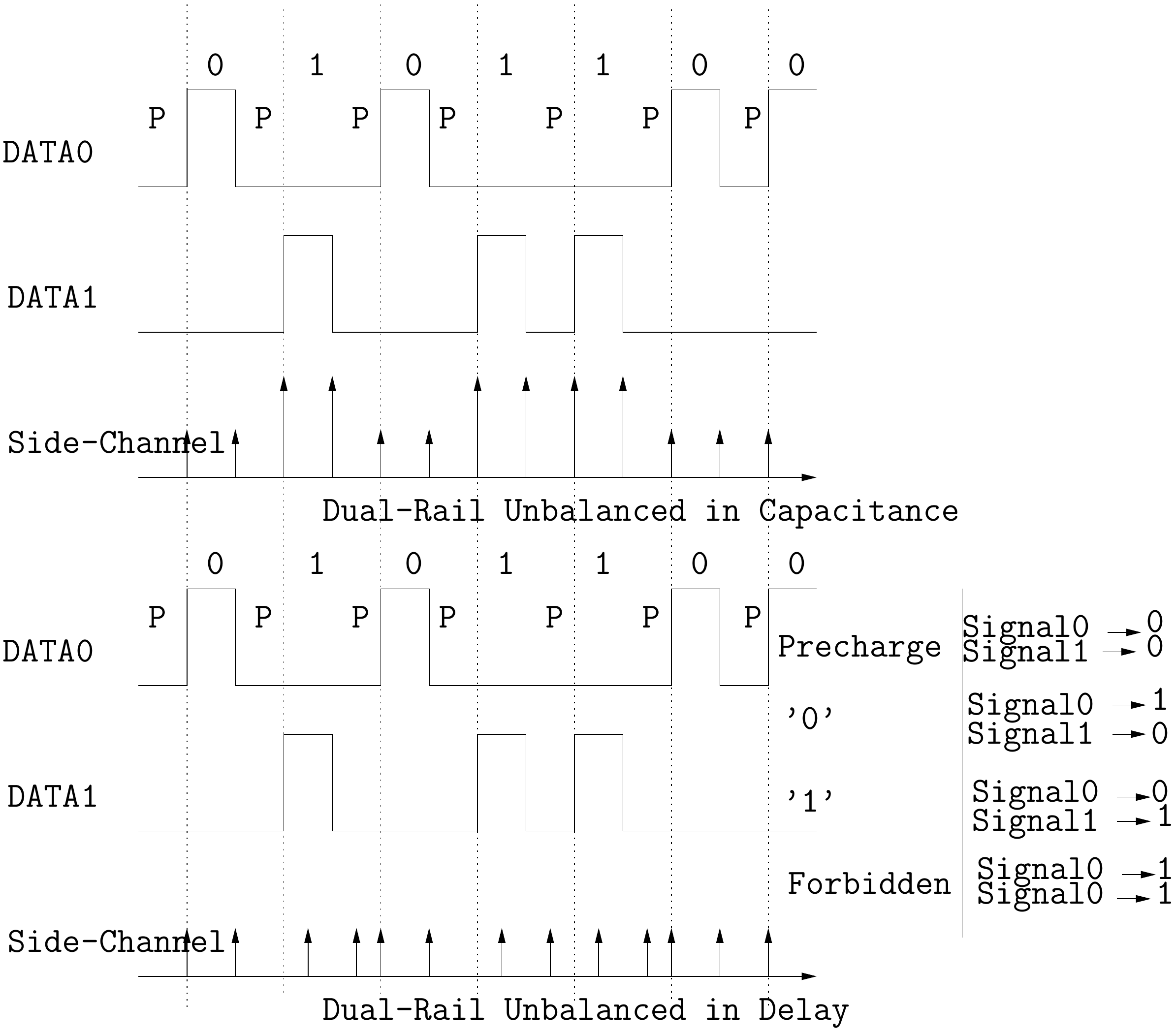}
\caption{Imbalance in capacitance and delay leakage information.}
\label{fig:imbalance}
\end{figure}
Figure~\ref{fig:imbalance} shows the effect of imbalance in capacitance and delay in dual-rails, on side channel leakage. 
Even these small differences could be exploited for cryptanalysis~\cite{tiri-dac07}. In section~\ref{subsec:Dynamic Power consumtion model} we have modelled each 
interconnect by its positive and negative step current response.
Ideally:
\begin{itemize}
\item
the +ve and -ve step current response for each dual-rail should be identical.
\item
for a buffered net, the buffers used should be identical, and each segment between buffers should also have identical step current response.
\end{itemize}

As a measure of indiscernability, we use the cross-correlation of the step responses of each net. The higher the cross-correlation, the more difficult it is to predict which net has undergone transitions.

However to simplify the design procedure we make the following assumptions:
the same length of wire of same width, charging the same capacitances, has a similar step current response, that is consumes the same current and causes the same delay irrespective of any bends.

In this respect we also define equitemporal lines for $n$-wire signals.
An equitemporal line
(\raisebox{+0.1mm}{\includegraphics[width=12mm]{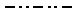}}\!)
is the set of points attainable simultaneously by signals originating from synchronized sources (\emph{i.e.} wave fronts.)

\subsubsection{\textbf{Indiscernability in EM emission}}
The radiation pattern measured at any point in space should be the same for each wire of a $n$-wire bus.
For example, given a set of parallel wires (not twisted), we can choose a point closer to one of the wires and further from others.
At that very point in space, radiation patterns emitted from different wires will be distinguishable.

At the gate level, the dual gates should be placed as close as possible, and at the net level we propose to route the dual-rail signals as a twisted pair (abridged ``T-Pair'') to deter EMA.

\section{Logic Block Architecture}
\label{sec:Logic Block Architecture}


\begin{figure}
\centering
{
\subfigure[PLB Functional Overview]{\includegraphics[width=0.24\textwidth]{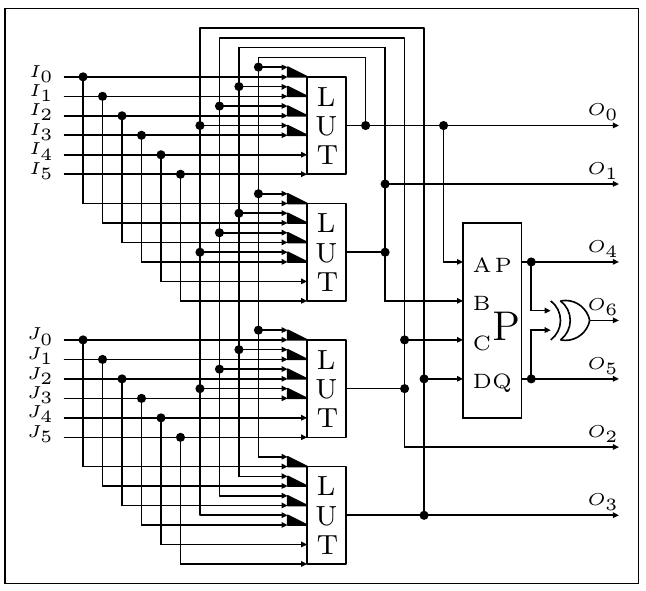}
\label{07-rec-global}}
\subfigure[Block P in Detail]{\includegraphics[width=0.24\textwidth]{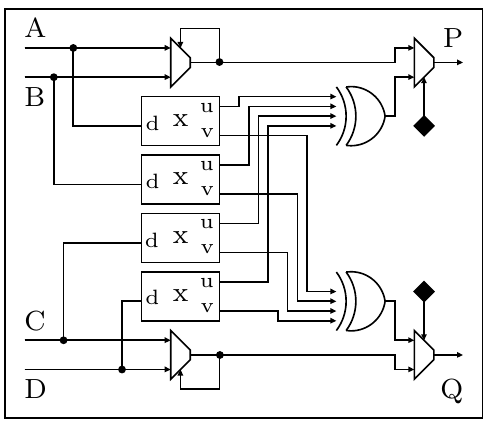}
\label{07-rec-blocp}}
\subfigure[Block X in Detail]{\includegraphics[width=0.25\textwidth]{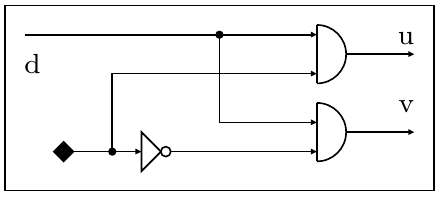}
\label{07-rec-blocx}}
}
\caption{PLB overview.}
\label{fig:PLB}
\end{figure}

Fig.~\ref{07-rec-global} depicts the structure of the PLB (Programmable Logic Block) able to handle the main QDI asynchronous styles. Black triangles at the left of LUTs are multiplexers controlled by programming points.
The P output block is described more precisely by Fig.~\ref{07-rec-blocp} and Fig.~\ref{07-rec-blocx}, in which the
black diamonds represent programming points. The feedbacks from output to inputs are used to obtain the memory effect.
%

%
%
%

Details of the PLB and mappings can be found in~\cite{recosoc}. A summary of all the implementable styles in the proposed PLB is given in table~\ref{tab2}.
The results are given in number of PLBs.
The three configurations considered in this table are:
\\\hspace*{1ex} A: Dual Rail,   2-input gate with Acknowledge input,
\\\hspace*{1ex} B: Dual Rail,   3-input gate with Acknowledge input,
\\\hspace*{1ex} C: Triple Rail, 2-input gate with Acknowledge input.

\begin{table}
\caption{Mapping Capacity of different protocols}
\label{tab2}
\begin{center}
\setlength{\extrarowheight}{2pt}
\begin{tabular}{cccc}
\cline{2-3}
\multicolumn{1}{c|}{} & A & B & C \\
\hline
2-phase EDGE  &   1 & n.a. & n.a. \\
2-phase LEDR  & 0.5 &    1 & n.a. \\
4-phase       & 0.5 &    1 &    2 \\ 
\hline
\end{tabular}
\end{center}
\end{table}

For the purpose of this article, we explain mapping of $1$-out-of-$2$, 2-input Gates onto the PLB here. Other mappings can be found in~\cite{recosoc}.

\subsubsection{1-out-of-2, 2-Input Gates}
\label{sec:1-out-of-2, 2-Input Gates}
\begin{figure}[t]
\centering
\includegraphics[width=0.25\textwidth]{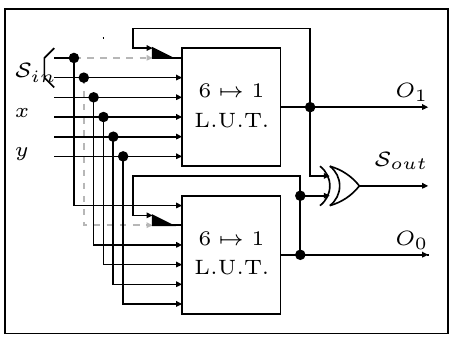}
\caption{Mapping of 1-out-of-2, 2-input Gates onto the PLB}
\label{fig:mapping}
\end{figure}

Let $f(x,y):\gf\times\gf\mapsto\gf$ a two-variable Boolean function. The
inputs are represented by 4 wires: $x_0$, $x_1$, $y_0$ and $y_1$, to which
a synchronization signal (\textit{Acknowledge}), \ACKIN, is added and the output signal $O$
represented by two wires: $O_0$ and $O_1$, together with an acknowledge output \ACKOUT.
The individual values of these wires are functions of $x$ and $y$, respectively denoted $f^0(x,y)$ and $f^1(x,y)$.

The equations of the outputs are:
\begin{equation}
	\begin{array}{l@{\,=\,}l}
		O_1 &
			\begin{cases}
				f^1(x,y) & \text{if } x,y\not=\INV \wedge \ACKIN = 0 \,,\\
				0        & \text{if } x,y =\INV \wedge \ACKIN = 1    \,,\\
				O_1      & \text{otherwise}                          \,,\\
			\end{cases}
		\\
		O_0 &
			\begin{cases}
				f^0(x,y) & \text{if } x,y\not=\INV \wedge \ACKIN = 0 \,,\\
				0        & \text{if } x,y =\INV \wedge \ACKIN = 1    \,,\\
				O_0      & \text{otherwise}                          \,,\\
			\end{cases}
	\end{array}
	\label{eq-4-ph-2-in}
\end{equation}
\noindent in which ``$x,y=\INV$'' stands for ``$x=\INV\wedge y=\INV$''.

Eq.~\eqref{eq-4-ph-2-in} shows that $O_0$ and $O_1$ are functions of 6 Boolean variables. 
Thus the minimal practical size for the LUTs is 64 bits, which can implement a 6-bit $\mapsto$ 1-bit function. 
As there are two output bits the minimal size of the PLB is 2 LUTs. 
The \ACKOUT output can be computed as $(O_1 \vee O_0)$. 
However, for homogeneity with the 2-phase protocols, we use $(O_1 \oplus O_0)$ instead. 
Note that, as the $O_0=O_1=1$ state is forbidden, the two functions are identical on the allowed domain.

A feedback is necessary to obtain the memory effect. 
One of the inputs to the LUTs must thus be programmable between the input to the PLB and the feedback.
As the inputs to the LUTs are the same, with the exception of the feedback wires, there can be a single connection box to the routing network.
Fig.~\ref{fig:mapping} shows the minimal structure of the PLB, which allows to implement 2-input gates with synchronization.

The associated wiring is:
\begin{equation}
\begin{array}{l@{\,=\,\text{LUT6}(\,}l@{\,,\,}l@{\,,\,}l@{\,,\,}l@{\,,\,}l@{\,,\,}l@{\,)}c}
O_0 & O_0 & \ACKIN & x_1 & x_0 & y_1 & y_0 & \,,\\
O_1 & \ACKIN & O_1 & x_1 & x_0 & y_1 & y_0 & \,.\\
\end{array}
\label{4-ph-2-in-prog}
\end{equation}

In Eq.~\eqref{4-ph-2-in-prog}, each wires of $x$ and $y$ is loaded with exactly the same number of inputs.
Note that the \ACKIN signal is connected twice to the routing network.

\subsubsection{Balanced LUT Implementation}
\label{sec:Balanced LUT Implementation}

\begin{figure}[t]
\centering
\includegraphics[width=0.5\textwidth]{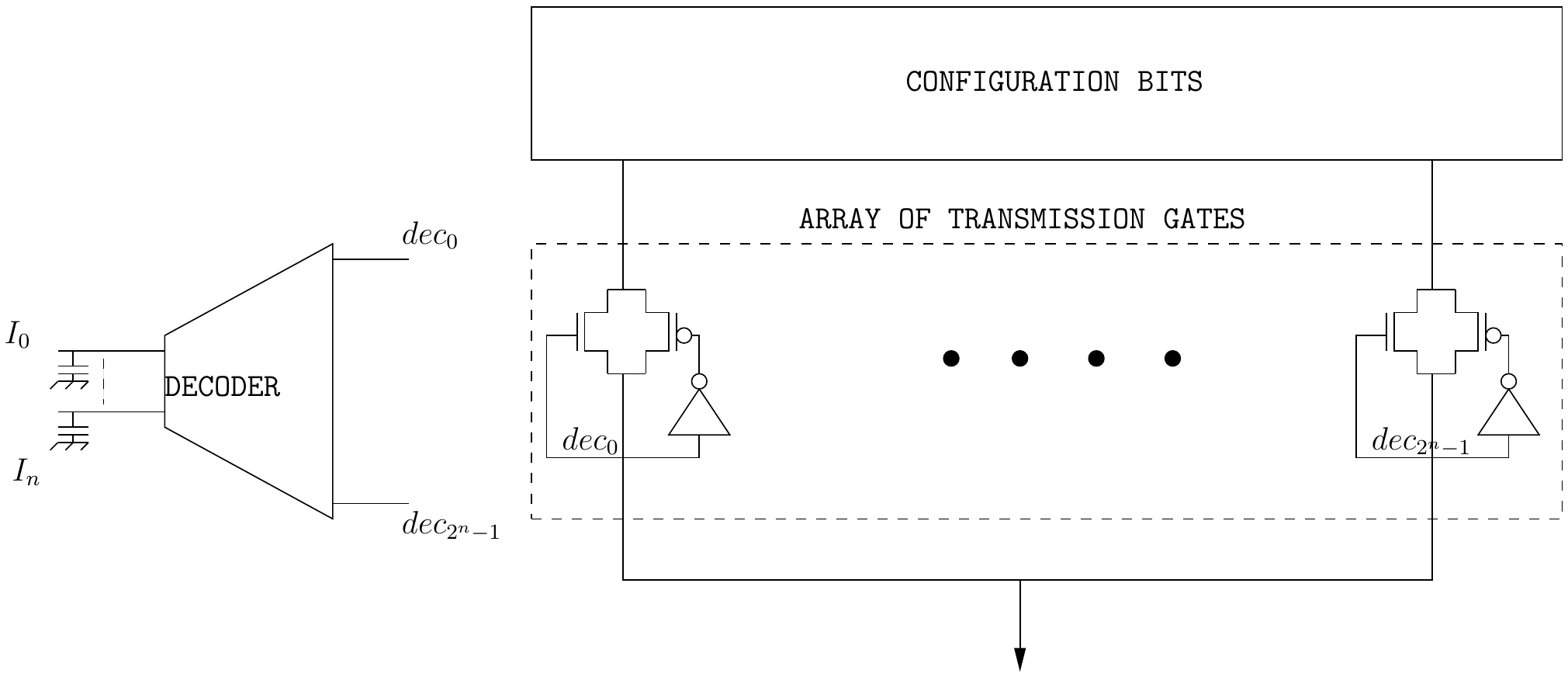}
\caption{LUT implementation with a wired AND.}
\label{fig:lut_impl}
\end{figure}

Figure~\ref{fig:lut_impl} depicts the LUT implementation scheme to achieve the objectives set in section~\ref{sec:Secure Place-Route Objectives}. In a classical LUT implementation with multiplexer trees each input is loaded with a different capacitance, and also the logic depth from configuration bits to the outputs varies with the input activity.

To circumvent these drawbacks, all input capacitances have been balanced by correctly adjusting the sizes of decoder's inverter, and a unique logical depth is imposed between the configuration bits and the LUT output. More details about the LUT implementation and simulation results can be found in \cite{icfpt}.

\begin{figure}[t]
\centering
\includegraphics[width=0.5\textwidth]{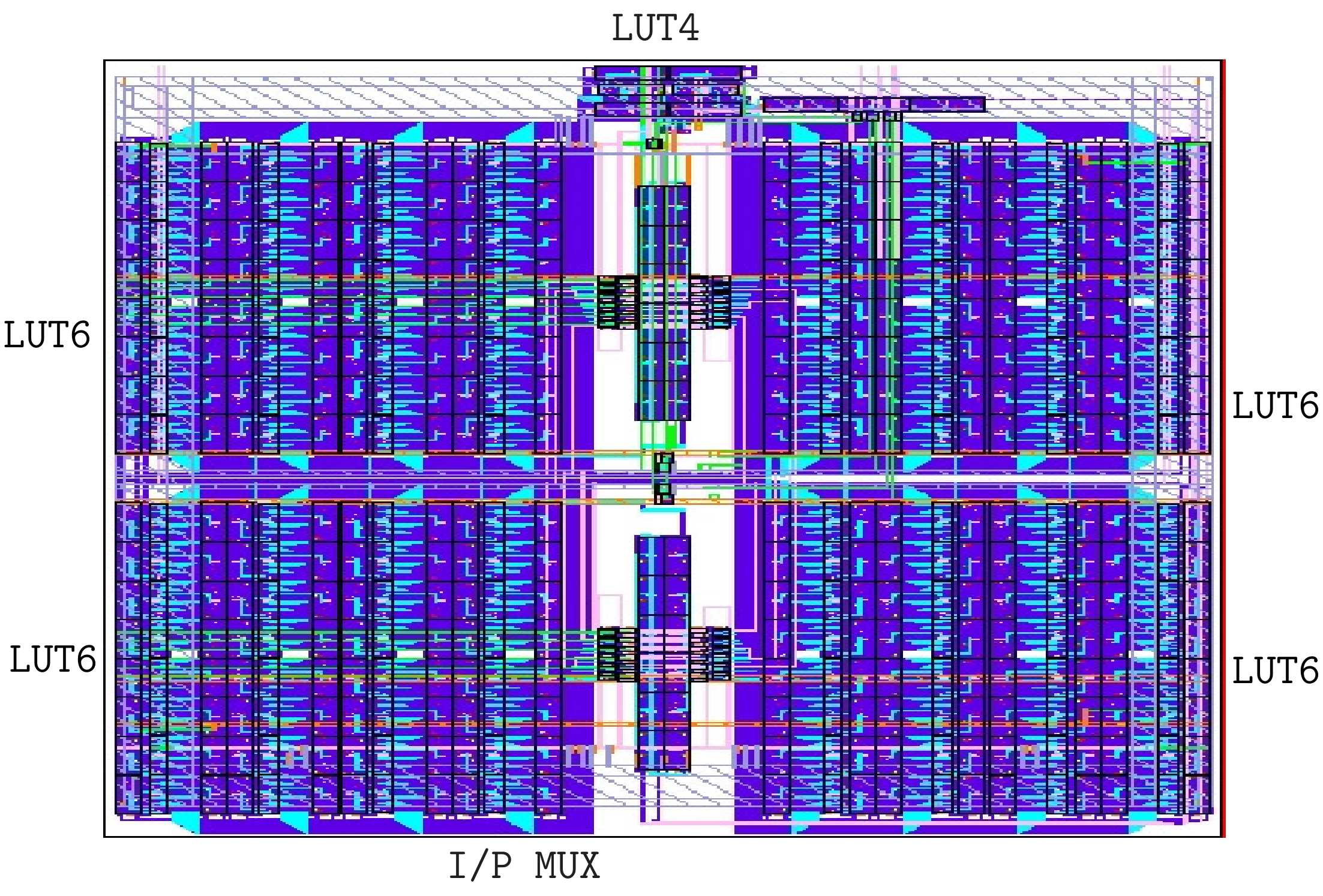}
\caption{PLB layout in Cadence \textsc{Virtuoso}.}
\label{fig:PLB_layout}
\end{figure}

Figure~\ref{fig:PLB_layout} describes the actual PLB layout. The four $6\rightarrow1$ LUTs are placed symmetrically, and in between the input multiplexers are placed
which connects the LUT inputs to the routing resources, and feedbacks. The block P is placed at the top. All the twelve inputs are placed at the top, and the seven outputs at the right of the  PLB layout.

\section{Routing Architecture}
\label{sec:Routing Architecture}

We use the traditional mesh routing architecture and associated nomenclature as explained in~\cite{vpr_book}.

\subsection{Subset Routing Architecture}
\begin{figure}
   \centering
   {\subfigure[\emph{Subset} switchbox using six-way switch-points.]{
   \includegraphics[width=0.43\textwidth]{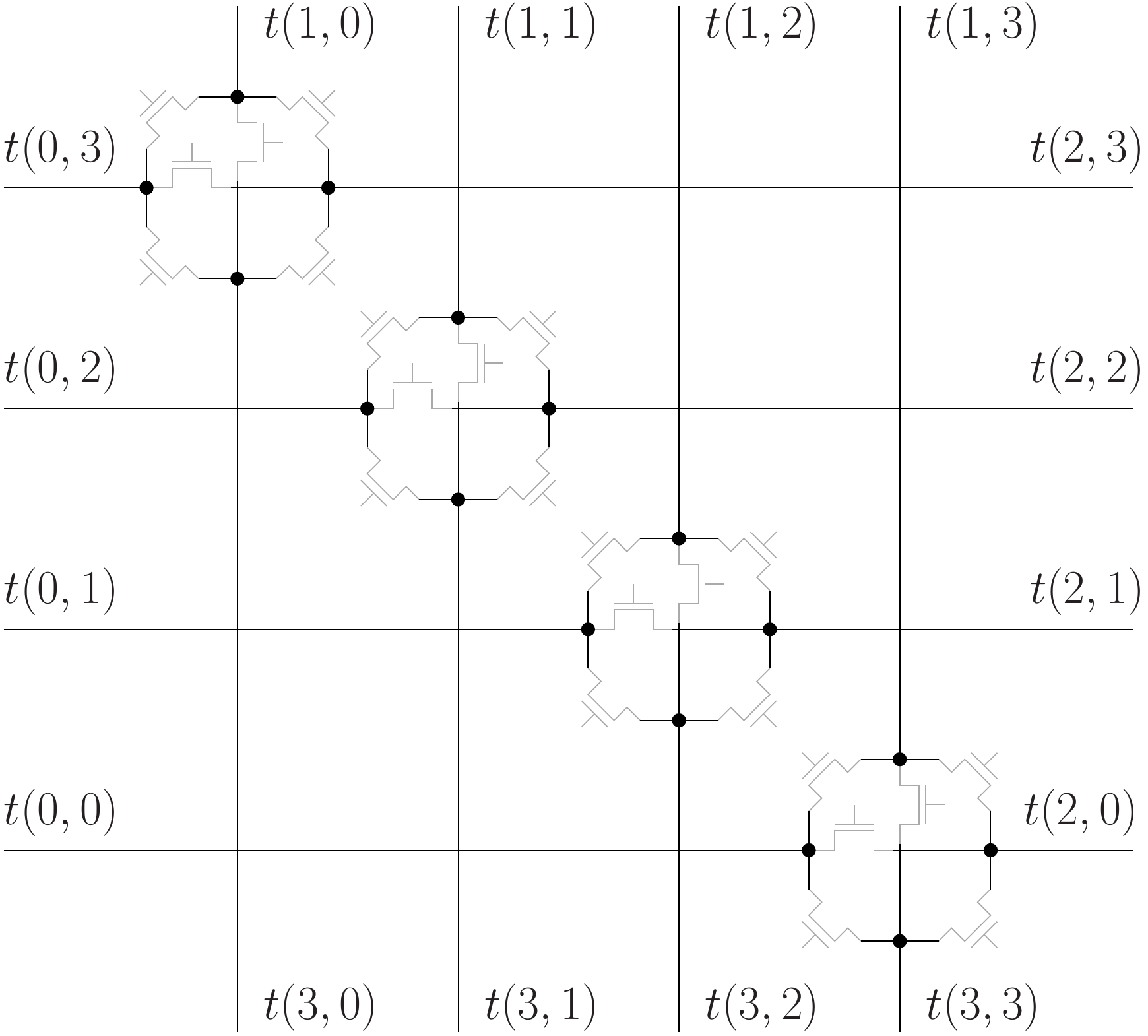}
   \label{fig:switch_matrix_subset}}
   \hspace{10mm}
   \subfigure[Equitemporal lines for \textit{subset} switchbox routing.]{
   \includegraphics[width=0.43\textwidth]{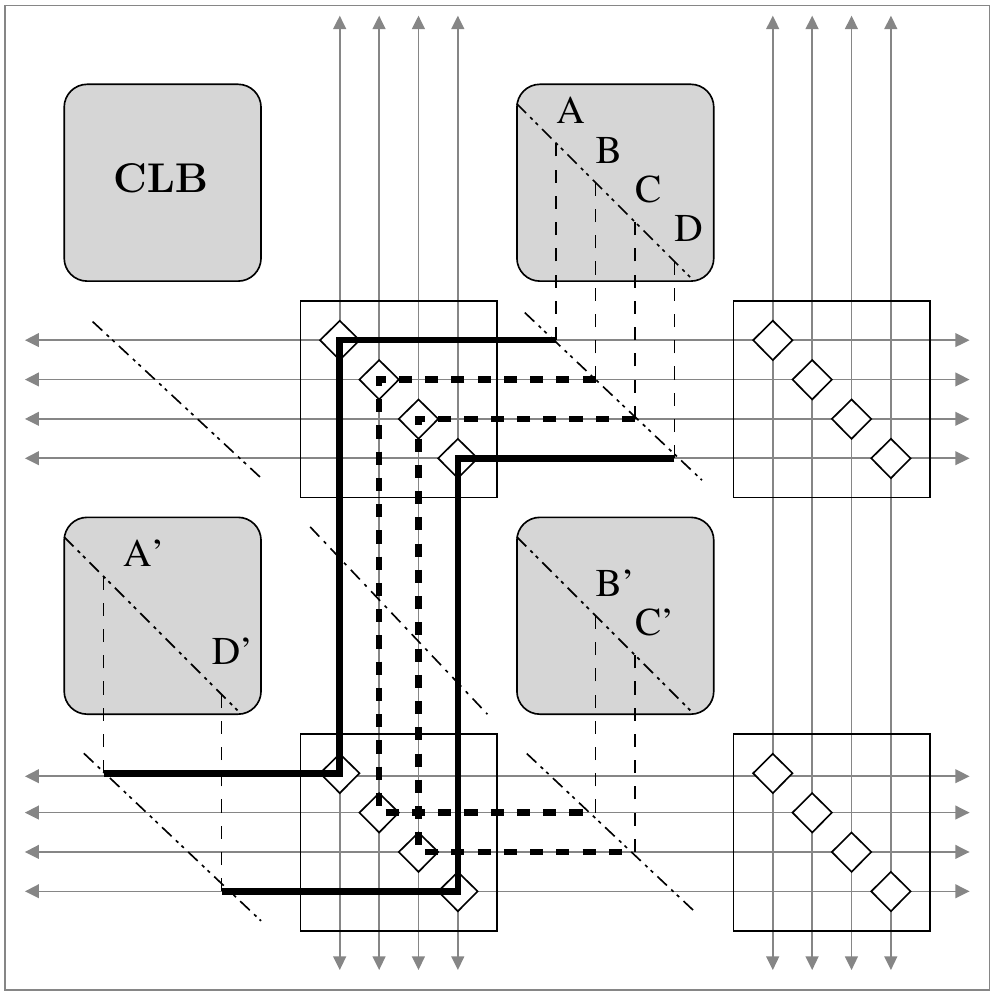}
   \label{fig:subset}}}
   \caption{\textit{Subset} switchbox Routing.}
\end{figure}

A subset switchbox~\cite{vpr_book} can be built by repeating a basic six-way switch-points along a diagonal, as shown in figure~\ref{fig:switch_matrix_subset}.
We consider that the diagonal formed by the six-way switch-points makes up equitemporal signals (see section~\ref{sec:Secure Place-Route Objectives}) if these signals are outputs of the same FPGA logic element CLB. 
Figure~\ref{fig:subset} shows the routing matrix using a subset switchbox.
Connection boxes from the equitemporal lines to the CLB inputs/outputs are considered as equitemporals.
They are discussed in section~\ref{subsec:connect}.
In figure~\ref{fig:subset}, the dual pair signals corresponding to connections \{A, A'\} and \{D, D'\} have exactly the same length and the same electrical characteristics.
The same goes for buses \{B, B'\} and \{C, C'\}.
Notice that the dual-rail signals are not necessarily routed in an adjacent way (case of A and D) and that it is possible to route in the same fashion multi-wire signals.

\subsection{Twisted Pair Routing Architecture}

\begin{figure}[t]
   \centering
   {\subfigure[Electric \& magnetic fields orientation in an un-twisted (a) and in a twisted (b) pair.]{
   \includegraphics[width=0.5\textwidth]{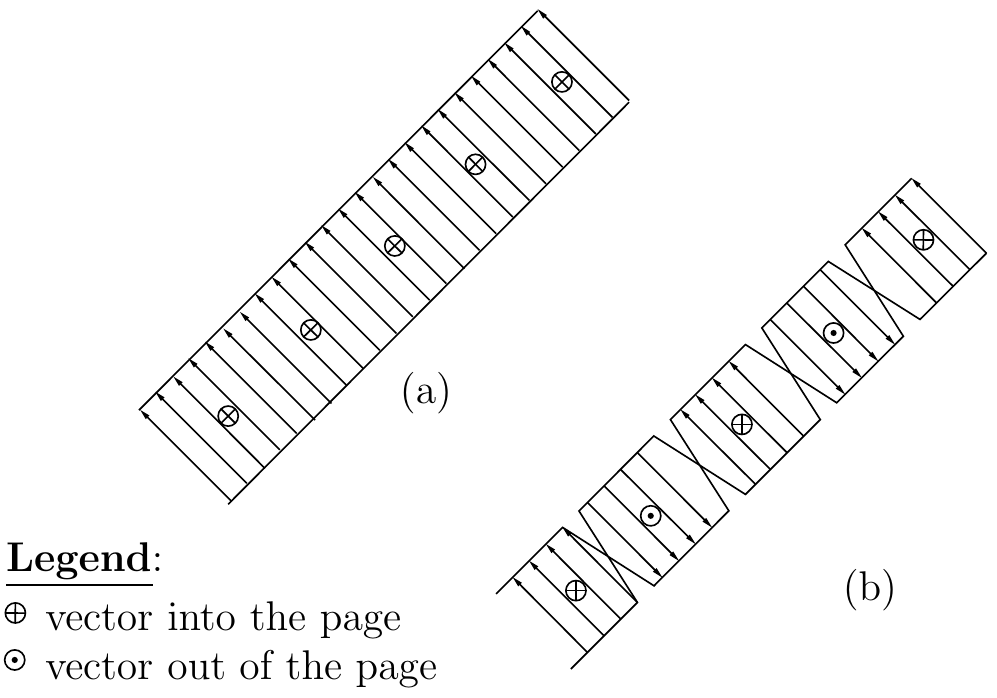}
   \label{fig:radiation}}
   \hspace{1cm}
   \subfigure[Equitemporal lines for the \textit{twisted-pair} switchbox.]{
   \includegraphics[width=0.35\textwidth]{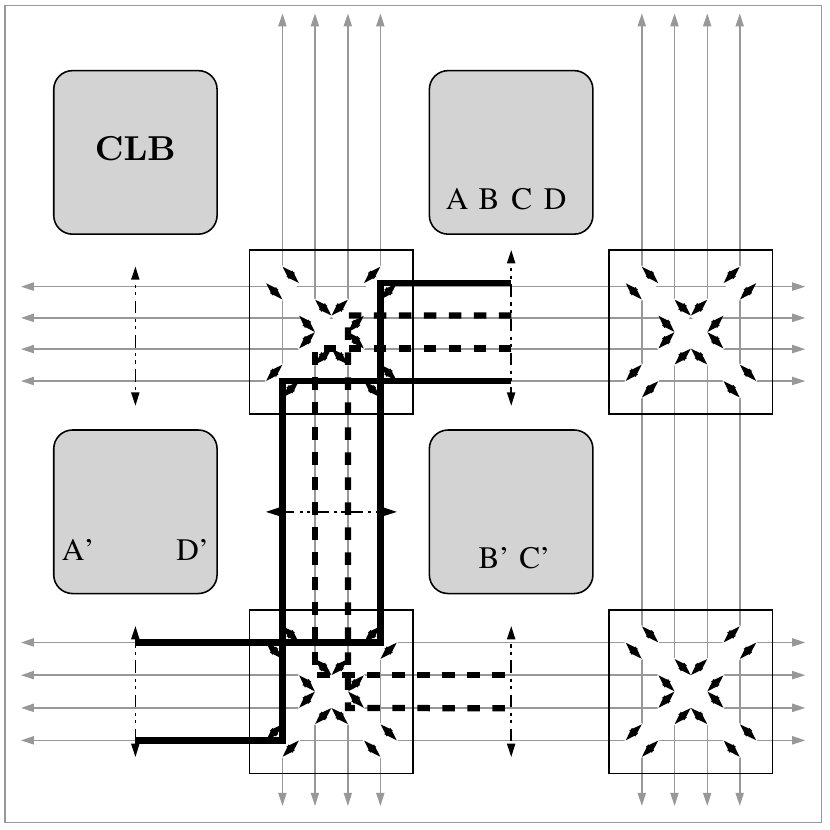}
   \label{fig:twisted_pair}}}
   \caption{The \textit{twisted-pair} switchbox.}
\end{figure}

As a countermeasure against information leakage through EM radiations, we propose to route every $n$-rail signal as a twisted bus.
Figure~\ref{fig:radiation} shows the advantages of using a twisted pair compared to parallel routed wires.
If we consider the twisted pair as made up of several elementary radiating loops, we see that the radiation from a loop is cancelled by that of adjacent loops.

In addition to reducing EM compromising radiations (outputs),
the twisted bus gains immunity from its EM vicinity (inputs).
Consequently, twisting signals bundles reduces cross-talk,

In order to route any $n$-rail signal as a twisted bus throughout the FPGA, two novel switchboxes are introduced in §\ref{sss-tot} and \ref{sss-ta}.

\subsubsection{Twist-on-Turn Switch Matrix}
\label{sss-tot}

The basic idea behind this switchbox is that every pair or $n$-uplets of signals deflected by the switchbox must come out twisted.
As shown in figure~\ref{fig:twisted_pair}, every $\pm \pi/2$ bend through this switchbox is a twisted pair.
We can express this switchbox using the notation described in~\cite{wilton} as:

{\tiny
\begin{displaymath}
\mathbf{S} =
	\bigcup_{i=0}^{W-1} \left\{ \begin{array}{c}
	\left[t(0,i),t(2,i)\right],\\
	\left[t(1,i),t(3,i)\right],\\
	\left[t(0,i),t(1,i)\right],\\
	\left[t(1,i),t(2,W-i-1)\right],\\
	\left[t(2,i),t(3,i)\right],\\
	\left[t(3,i),t(0,W-i-1)\right].
	\end{array}\right\}
\end{displaymath}
}

where each terminal is represented as \TERM{j}{i},
where $j$ denotes each subset corresponding to each side
(%
	0\,=\,left, %
	1\,=\,top, %
	2\,=\,right, %
	3\,=\,bottom%
)
and $i \in [0, W[$ denotes the position of the terminal in that subset.
Connection boxes from the equitemporal lines to the CLB inputs/outputs are considered as being equitemporal perpendicular to the routing channel.
They are discussed in section~\ref{subsec:connect}.
In figure~\ref{fig:twisted_pair}, the dual pair signals corresponding to connections \{A, A'\} and \{D, D'\} have exactly the same length even if they cross at the switching box.
It is exactly the same for buses \{B, B'\} and \{C, C'\}.

When turning, this switch matrix introduces a small imbalance for the arrival time on the deflecting switch point.
If the switch point is implemented  with \emph{passive} gates, this balance violation is not observable by an attacker.
The counterpart is that the channels must be buffered,
which can safely be done with \emph{active} gates, because every wire in a channel is equitemporal.

\subsection{Twist-Always Switch Matrix}
\label{sss-ta}

\begin{figure}
   {
   \subfigure[Twist-on-Turn.]{\includegraphics[width=0.21\textwidth]{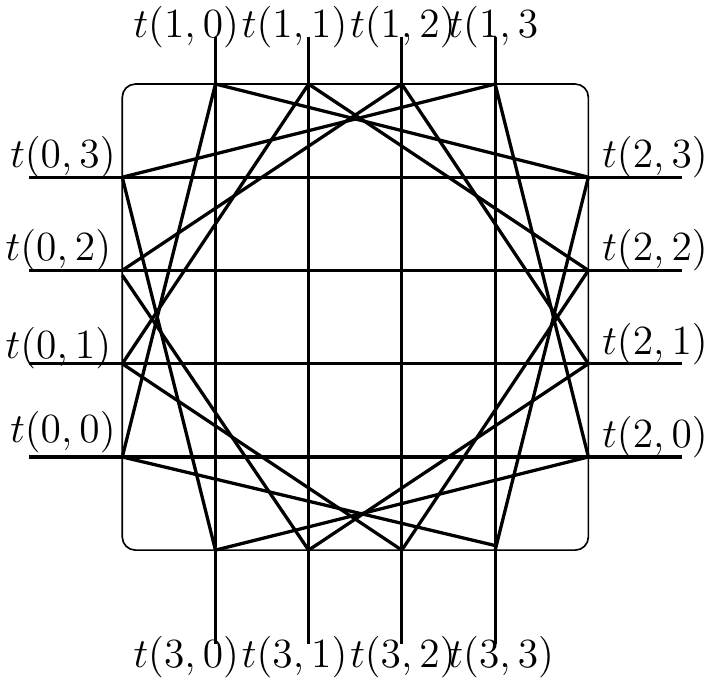}
   \label{fig:behave1}}
   \subfigure[Twist-Always.]{\includegraphics[width=0.21\textwidth]{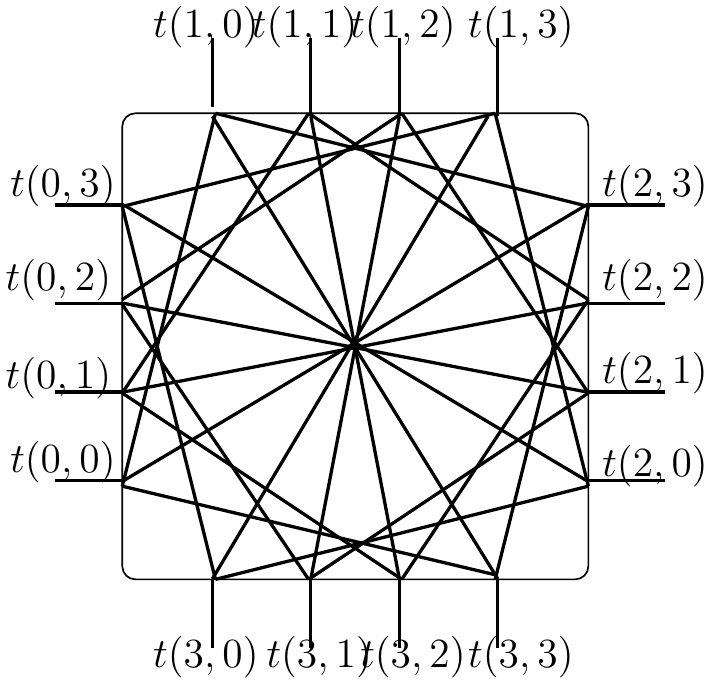}
   \label{fig:behave2}}

   \subfigure[Twist-always switch matrix layout scheme.]{
   \includegraphics[width=0.5\textwidth]{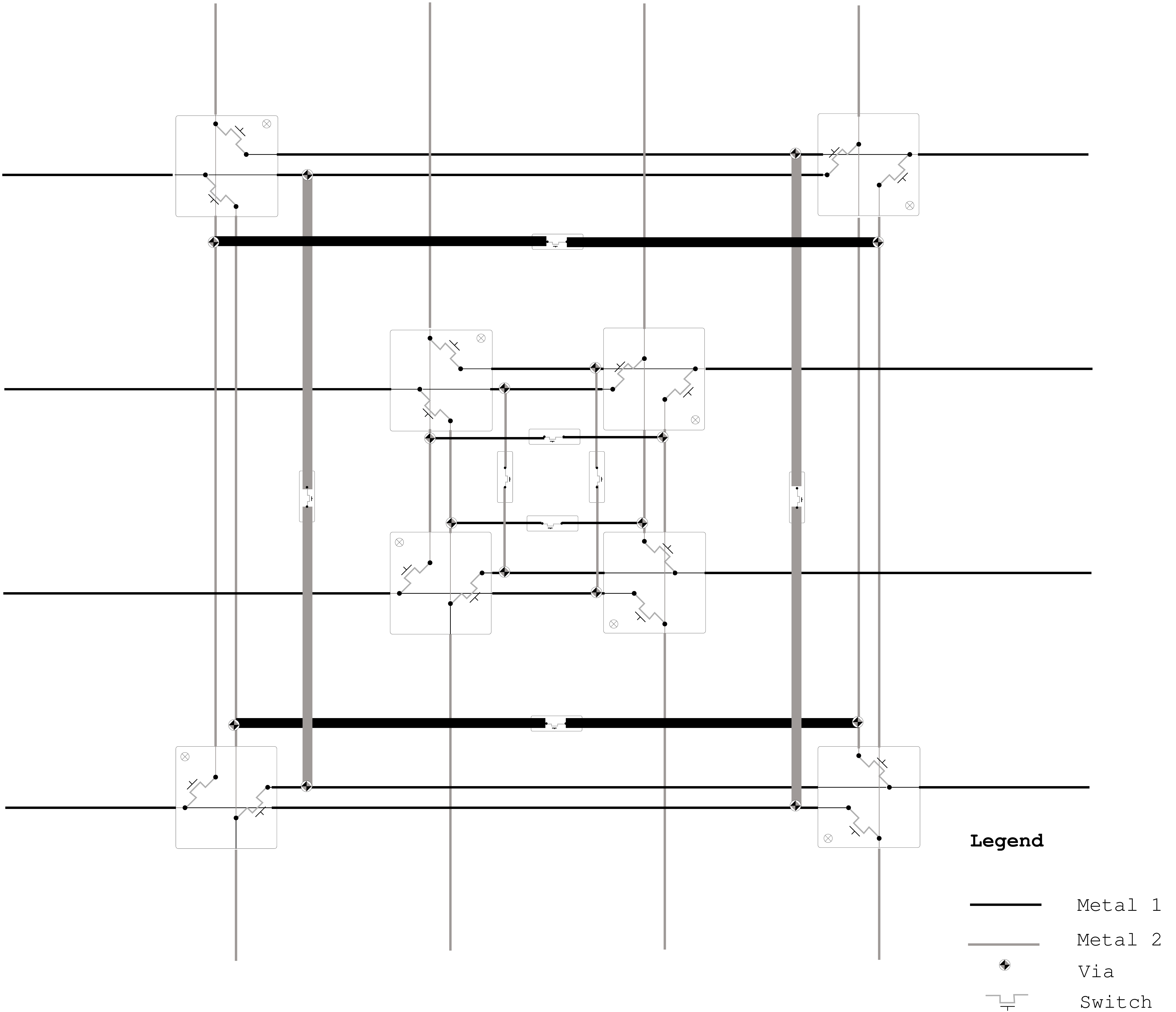}
   \label{fig:tpair}}
   }

   \caption{The twisted-pair switchboxes.}
   \label{fig:behave}
\end{figure}

The twist-on-turn matrix does not twist buses when they are routed straight.
This matrix can be transformed into a twist-always matrix by
twisting the wire $i$ with wire $W-1-i$ for straight connections, as shown in figure~\ref{fig:behave}, $W$ being the number of channels.

This matrix allows the use any 1-out-of-$n$ (asynchronous) style, as it is possible to twist a number of lines greater than two.
 
This switchbox cannot be implemented with traditional six-way switch-points, even if the number of transistors remains the same.
A possible implementation of the twist-always switch box is shown in figure~\ref{fig:tpair}.
It can be laid out in silicon with two interconnect layers and by repeating two basic patterns over space.
Note that for straight (\emph{e.g.} from left to right) connections, the outer rails are drawn wider than the inner rails to compensate for the difference in lengths.
Alternatively, every wire can keep the same nominal width, but inner rails are forced to zigzag so as to make up for their shorter length.
For bends, every rail traverses an equal distance, hence this compensation is not required.


These new switchboxes are close to conventional universal/subset switchboxes in terms of connectivity. Hence we can expect similar performance in routability of netlists in the FPGA.

\subsection{Connection Box Implementation}
\label{subsec:connect}

\subsubsection{Cross-Bar Connection Box}
As depicted in figures~\ref{fig:subset} and~\ref{fig:twisted_pair},
a signal routed from one equitemporal line to another has the same delay.
Therefore the connection box (C-Box) between the $W$ channel wires and the CLB $I \in [0, W[$ inputs/outputs should also keep this equitemporality.
We propose to use a crossbar connection box based on balanced binary trees,
built according to the following three rules:
\emph{(i)  } from the channel, $W$ trees have $I$ equal-length branches,
\emph{(ii) } from the CLB,     $I$ trees have $W$ equal-length branches,
\emph{(iii)} the two trees are superimposed orthogonally and the $W \times I$ branches from each tree type meet via a switch point.

Figure~\ref{fig:x-bar} illustrates the layout of the balanced crossbar with $W=4$ and $I=4$,
using only two metal layers (represented with two different thicknesses.)
The crossbar area is
$
W \cdot \lceil \log_2 \left( I \right) \rceil
\times                                       
I \cdot \lceil \log_2 \left( W \right) \rceil
$ square routing pitches,
and can be freely depopulated without altering its security level.


\subsubsection{C-Box for Subset \& Twisted-Pair Switch Matrix}
\begin{figure}[t]
   \centering
   {\subfigure[Balanced crossbar for the connection box.]{
   \includegraphics[width=0.3\textwidth]{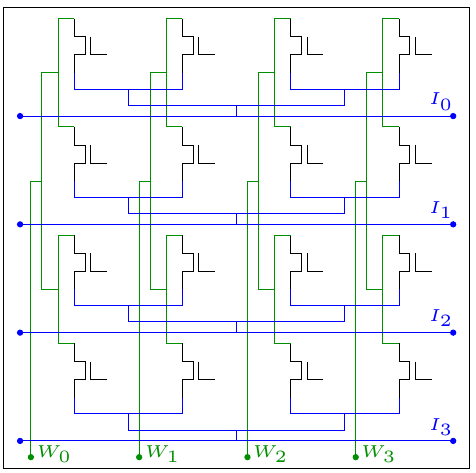}
   \label{fig:x-bar}}
	\subfigure[For Subset.]{\includegraphics[width=0.3\textwidth]{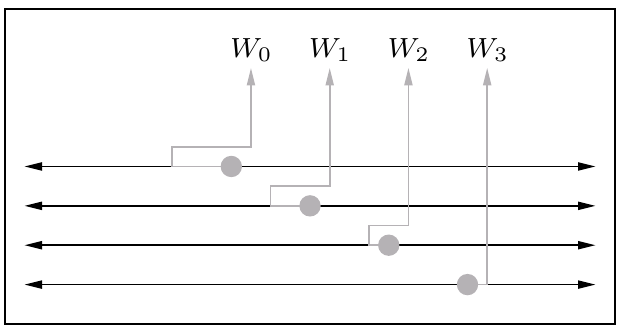}
	\label{fig:conn_subset}}
	\subfigure[For T-pair.] {\includegraphics[width=0.3\textwidth]{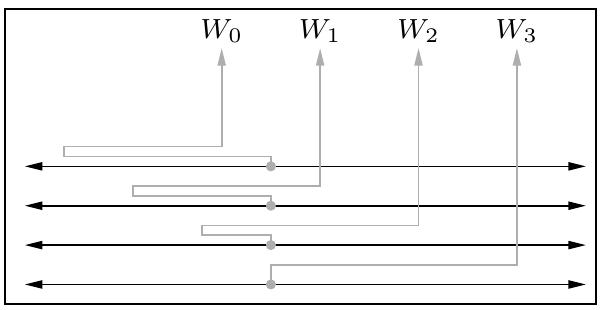}
	\label{fig:conn_tpair}}}
   \caption{Balanced crossbar and connections between channel wires and crossbar.}
   \label{fig:conn-channel}
\end{figure}

The equitemporal lines are either diagonal (for the subset switch matrix, \emph{cf} Fig.~\ref{fig:subset}) or horizontal/vertical (for the twisted switch matrix, \emph{cf} Fig.~\ref{fig:tpair}.)
The connections between the channel and the crossbar should compensate for the wire length delays.
A solution for both cases is illustrated in figure~\ref{fig:conn-channel}.

Example layouts and statistics of the T-pair switchbox, and the binary-tree connection box can be found in~\cite{chaudhuri-arc2008}.

\section{Single Driver Architecture}
\label{sec:Single Driver Architecture}

Single driver segments are shown to give better delay performances and better area-delay product for an FPGA~\cite{single_driver,vpr5}. These benefits are result of less loading
of interconnect segments, and availability of equal number of tracks in each direction as in traditional bidir-tri routing architecture.

\begin{figure}
\centering
{
\subfigure[Single Driver Subset Switch-Box]{\includegraphics[width=0.45\textwidth]{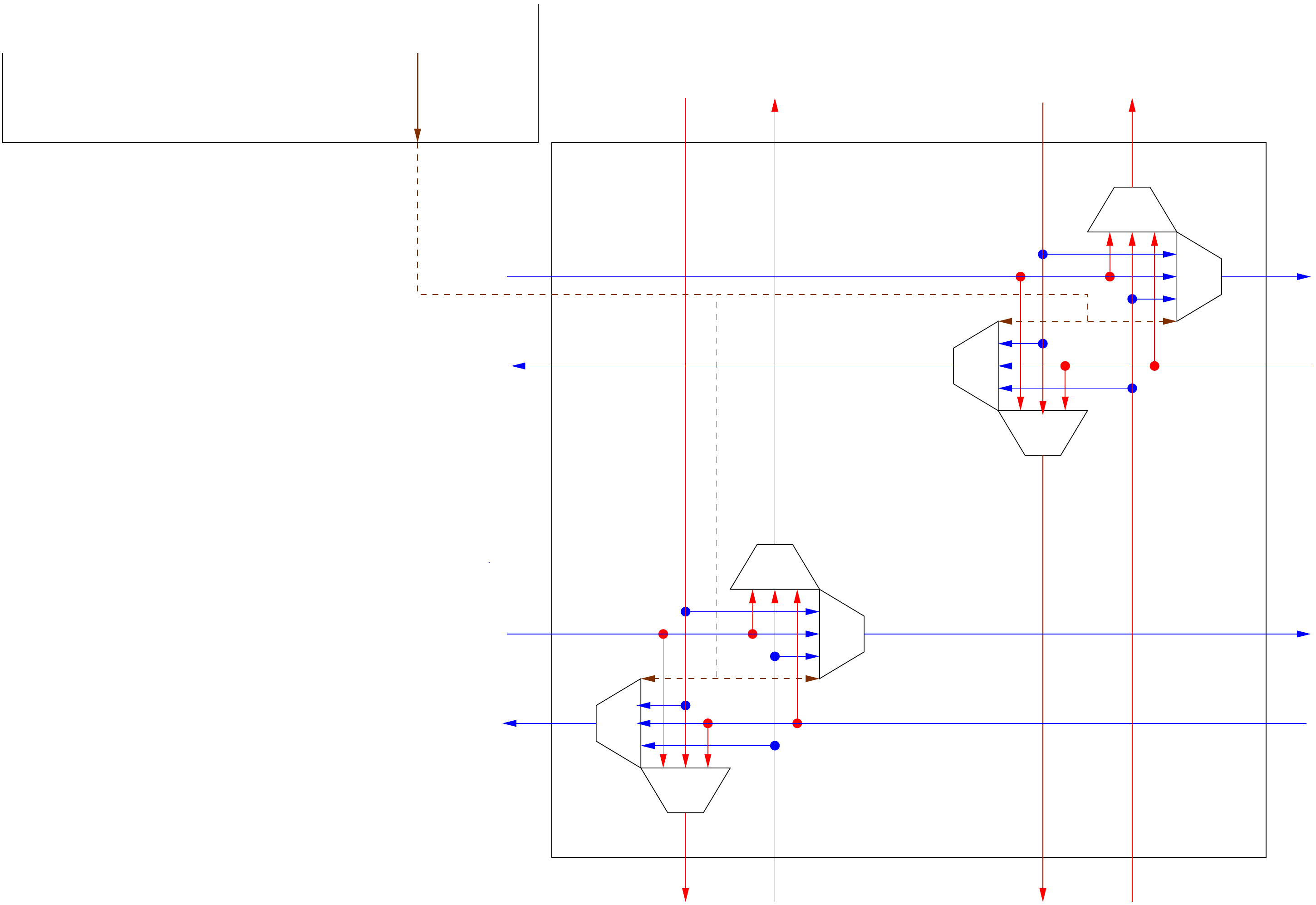}
\label{fig:subset_sd}}
\subfigure[Single Driver Tpair Switch-Box]{\includegraphics[width=0.45\textwidth]{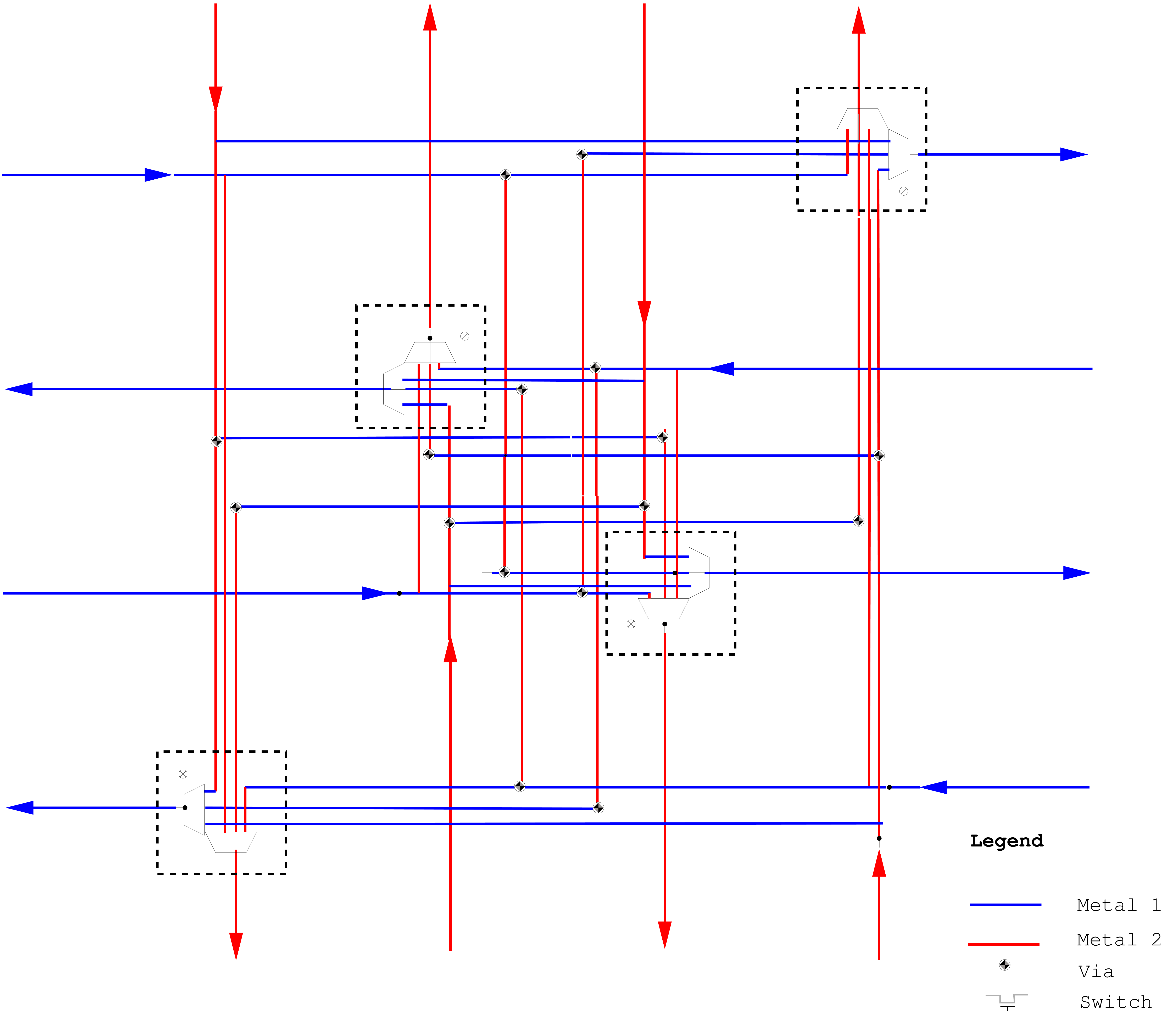}
\label{fig:tpair_sd}}
}
\caption{Porting the solution to single driver architecture.}
\label{fig:sd}
\end{figure}

Figures~\ref{fig:subset_sd} shows how the subset switchbox layout scheme~\ref{fig:switch_matrix_subset} can be ported to single driver architecture. Note that 
the connection box nets are routed as a binary tree in both X and Y direction.
Figure~\ref{fig:tpair_sd} shows the layout scheme for the T-pair switchbox with single driver interconnects. Note that this scheme too uses 
a basic switch-point as in~\ref{fig:tpair} which is rotated as required.

\section{Configuration Architecture}

\begin{figure*}[t]
\centering
\includegraphics[height=0.2\textheight,width=1.\textwidth]{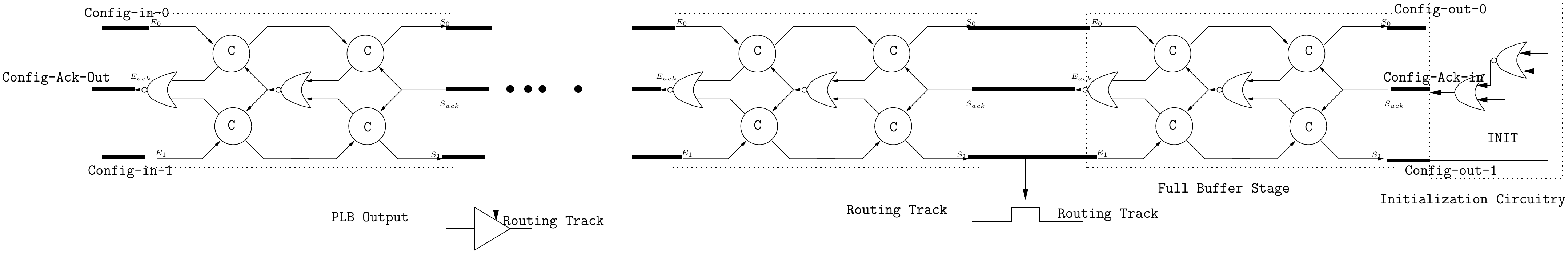}
\caption{The asynchronous configuration chain with reset circuitry.}
\label{fig:config_chain}
\end{figure*}
\begin{figure*}[t]
\centering
\includegraphics[height=0.2\textheight,width=1.\textwidth]{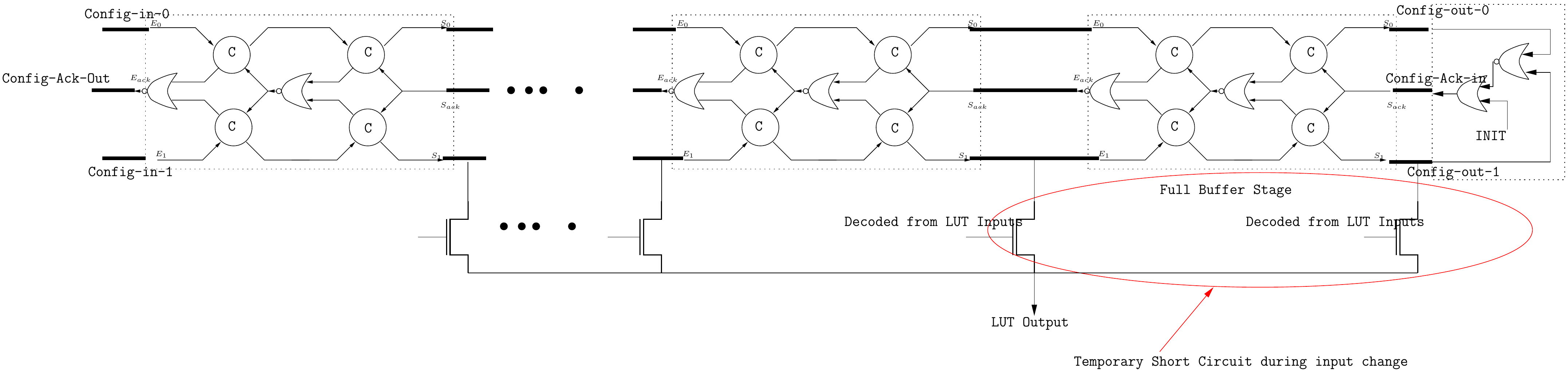}
\caption{If the LUT inputs are not defined, there could be a temporary short-circuit. Due to this bug the LUTs are not usable in the FPGA, need to be replaced by buffers instead of pass transistors.}
\label{fig:config_bug}
\end{figure*}

Figure~\ref{fig:config_chain} describes the configuration architecture for our asynchronous FPGA ``SAFE''.
The configuration chain is designed 
assuming $1$-out-of-$2$ $4$-phase protocol (as described in figure~\ref{fig:async2}). It consists of a series of Full Buffers
terminated by an \textit{initialisation circuitry}.
The function of the \textit{initialisation circuitry} is to bring the whole chain to (`0',`0') state, and to avoid any invalid state (`1',`1') after power up, or to erase a previous configuration. 
For \emph{initialisation} the signal \texttt{INIT} is put to `0', and (\texttt{config-in-0}, \texttt{config-in-1}) are held at `0'.
The `0's from the configuration input will propagate throughout the chain making a \emph{reset} action.
For any invalid state (`1',`1') after power up, we can see that the output of the \emph{nor} gates in the chain will be `0',
so the invalid state will be overwritten by (`0',`0') from the input of the chain. Since the signal \texttt{INIT} is held at '0' during the initialisation phase, the acknowledge 
input to the last stage is `0' for 
\begin{eqnarray*}
(\texttt{config-out-0}, \texttt{config-out-1}) & & \\
= (0,0) \text{ or } (0,1) \text{ or } (1,0) \,. & & 
\end{eqnarray*} 
It will only accept a (`0',`0') from the previous stage. 

During normal operation, \texttt{INIT} and consequently the acknowledge
input to the last stage is held at `1', hence the chain will accept any valid state until the chain is full.

\begin{figure}
   \centering
   {
   \subfigure[The initialisation and configuration phases.]{
   \includegraphics[width=0.5\textwidth]{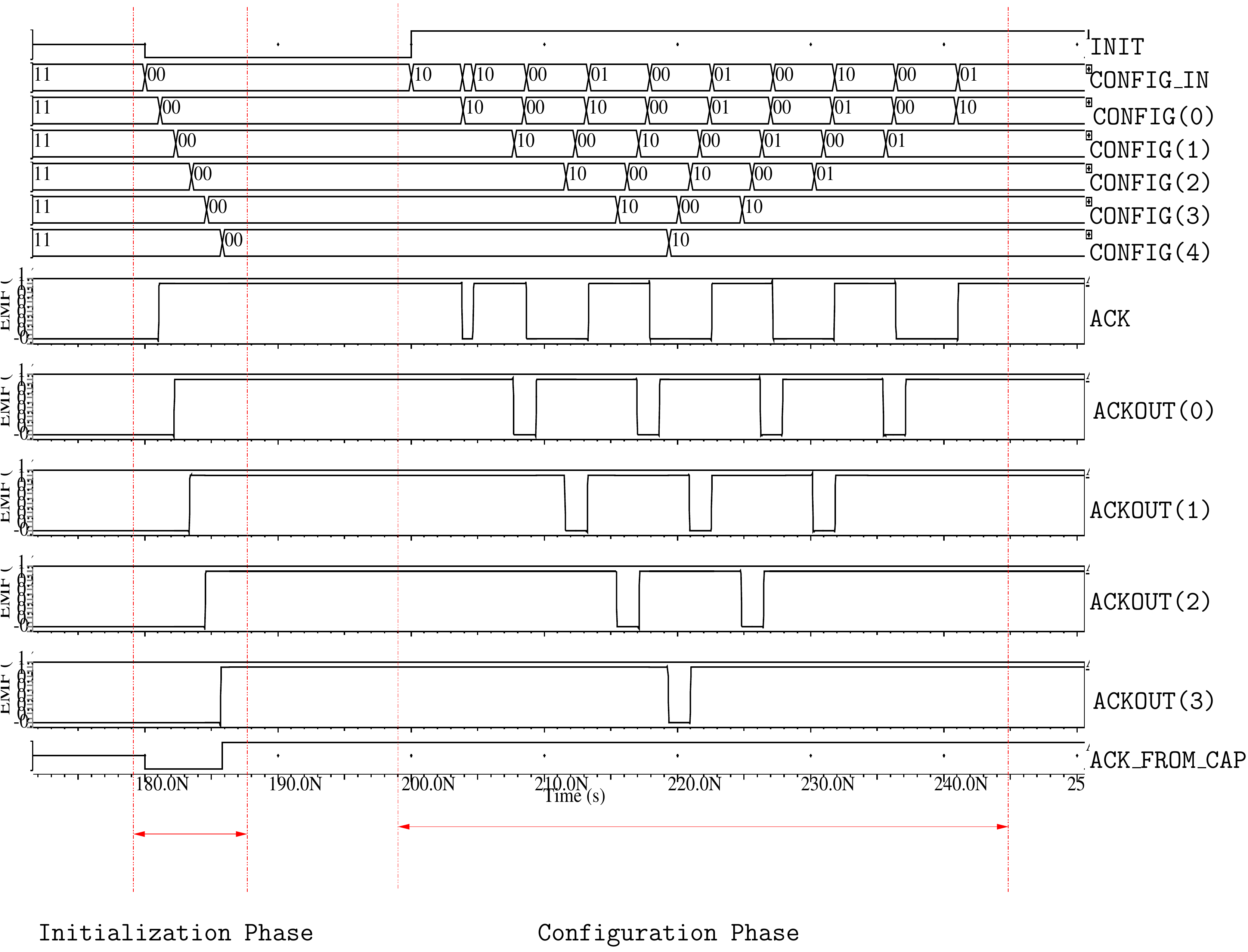}
   \label{fig:config}}
   \subfigure[The maximum speed of the Configuration Chain ($\sim$ 1.6~GHz), simulated with RC extracted view.]{
   \includegraphics[width=0.5\textwidth]{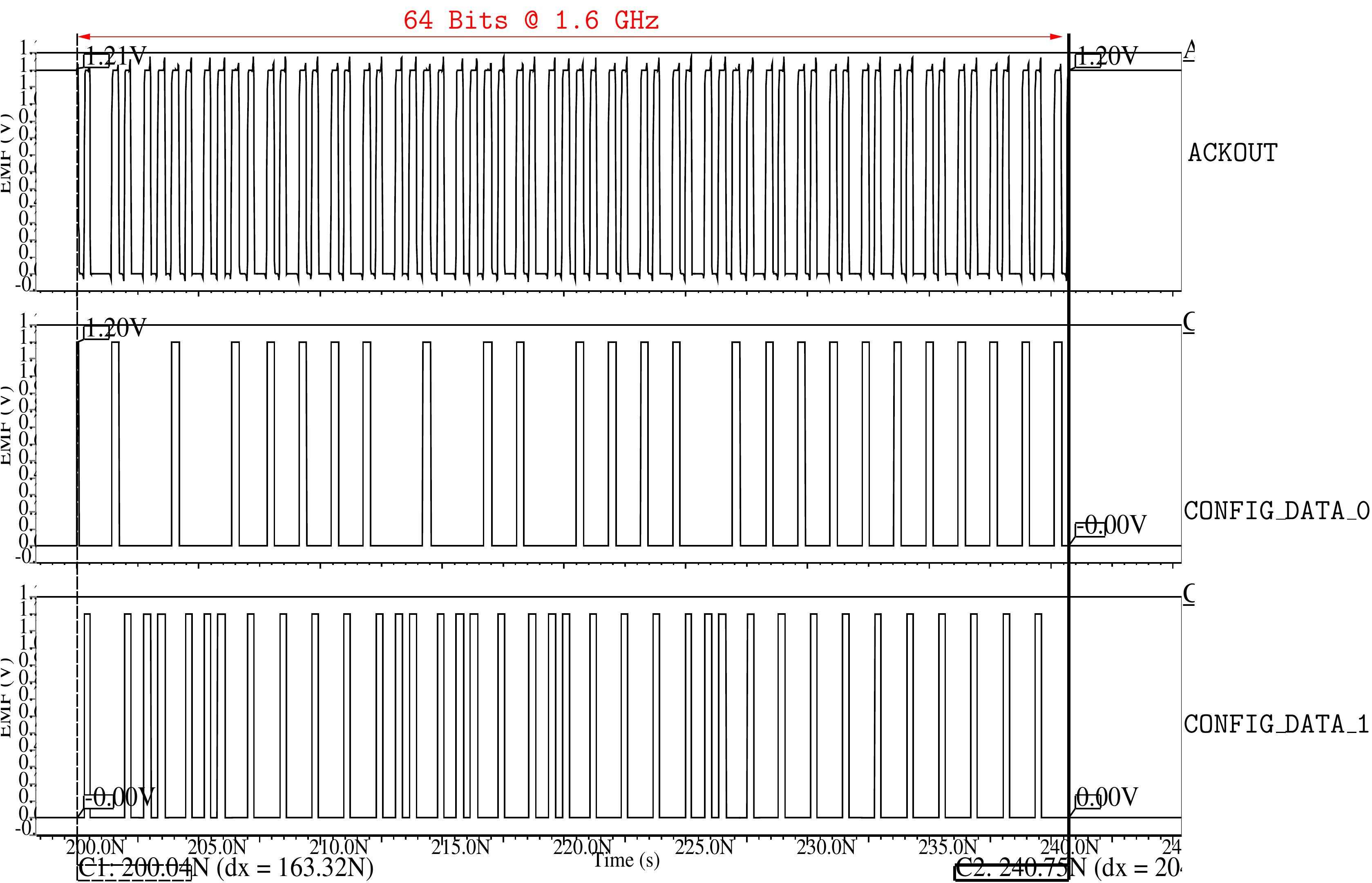}
   \label{fig:config_speed}}
}
   \caption{Simulation of the asynchronous configuration chain.}
\end{figure}

Figure~\ref{fig:config} shows the mixed-signal simulation results for an asynchronous configuration chain with five \emph{full-buffer} stages
and a initialisation cap as explained in fig.~\ref{fig:config_chain}. During the initialisation phase, the input to the configuration chain \texttt{CONFIG\_IN}
and the \texttt{INIT} signal is forced to `0'. We can see that the (`0',`0') propagates along the configuration chain, and the acknowledge signals are
initialised to `1' (READY) state.
We can see that even if the configuration signals are powered up at an invalid (`1',`1') state, the reset function of the
\emph{initialisation cap} brings the chain to a valid \emph{precharge} state which is ready to accept a new configuration.

In the configuration phase the \texttt{INIT} signal is kept at `1', and we can see in fig.~\ref{fig:config} the configuration bits advancing through the chain.
The waveforms presented are from a mixed-signal simulation, where the five \emph{full-buffer}s are analog, the \emph{initialisation cap} and the testbench controlling
\texttt{CONFIG\_IN} are digital circuits. Because of this reason the signals \texttt{ACK} and \texttt{ACK\_FORM\_CAP} at two ends of the chain have different delays.

In Figure.~\ref{fig:config_speed} we present the simulation results of the configuration of a single 6-input LUT with parasitic RC to
determine the highest achievable configuration speed. We used STARRCXT for parasitic extraction and ADVANCE MS (tool from Mentor Graphics~\cite{mentor}) for simulation. In the waveforms
the input is the digital input to the configuration chain and the acknowledgement (\texttt{ACK}) from the chain which is a analog circuit with parasitic RC.
The rise-time ($t_r$), fall-time ($t_f$) and delay of the virtual analog to digital converters for simulation are kept very small ($\sim0.1$ ps) so that they do not affect the
simulation results.

From fig.~\ref{fig:config_speed} we can see that we can configure 64 bits in about $\sim40$~ns. Thus the maximum achievable configuration speed
is around $\sim1.6$~GHz. This high speed is particular to the asynchronous configuration chain. This is mainly due to the fact that the each configuration
stage output see very small capacitive load, since it has a fanout of 2: one for the next stage and one for the switch connected to the output (see fig.~\ref{fig:config_chain}).
Although this is also true for synchronous shift register chains, their speed is often limited by the clock tree skew.
The reader might note that there is a small difference in the acknowledgement (\texttt{ACK}) delay for \texttt{CONFIG\_DATA\_0} and \texttt{CONFIG\_DATA\_1}. This is
due to the fact that in our FPGA the switches are connected only to \texttt{CONFIG\_DATA\_1}, thus it has a bigger delay than \texttt{CONFIG\_DATA\_0}. This is also particular to the
asynchronous configuration chain, because in the synchronous case we had to use the worst-case delay to design the clock tree.

\begin{table}
\begin{center}
\caption{No. of Total Configuration Bits.}
\label{tab:config_count}
\begin{tabular}{|l|p{0.5cm}|p{2cm}|p{1cm}|}
\hline
\texttt{SubModule}     & \texttt{Qty.} & \texttt{Switch Count} & \texttt{Total}\\ \hline
\texttt{PLB}       & \texttt{9} & \texttt{287}   & \texttt{2583}\\ \hline
\texttt{PLB Connection Box}       & \texttt{9} & \texttt{$(12+7)\times 8$} & \texttt{1368}\\ \hline
\texttt{IO Connection Box} & \texttt{12} & \texttt{$(3+3)\times 8 \times 0.5$} & \texttt{288}\\ \hline
\texttt{IO Config Bits} & \texttt{12} & \texttt{$3\times 12$} & \texttt{36}\\ \hline
\texttt{Switchbox(Full)} & \texttt{4} & \texttt{$6\times 8$} & \texttt{192}\\ \hline
\texttt{Switchbox($1/2$)} & \texttt{8} & \texttt{$3\times 8$} & \texttt{192}\\ \hline
\texttt{Switchbox($1/4$)} & \texttt{4} & \texttt{$1\times 8$} & \texttt{32}\\ \hline
\texttt{} & \texttt{} & \texttt{} & \texttt{4691}\\ \hline
\end{tabular}
\end{center}
\end{table}
Table~\ref{tab:config_count} explains the count of configuration bits in our prototype. Note that the following formulas are illustrated in
counting the switches of the routing ressources:
\begin{itemize}
\item
No. of switches in a switchbox with $W$ channels and $N$ sides = $\binom{N}{W} \times W$
\item
No. of switches in a connection box with $N_I$ inputs, $N_O$ outputs and $W$ channels =$ (N_I +N_O) \times W \times F_c$.
\end{itemize}
In our case $W=8$, for PLBs $N_I=12$ and $N_O=3$ and for IOBs $N_I=3$ and $N_O=3$.
For PLB connection boxes, $F_c=1$ but for IOB connection boxes $F_c=0.5$

\section{Prototype}
\label{sec:Prototype}

\begin{figure}
\centering
{
\subfigure[Chip Micro-Graph.]{\includegraphics[width=0.48\textwidth]{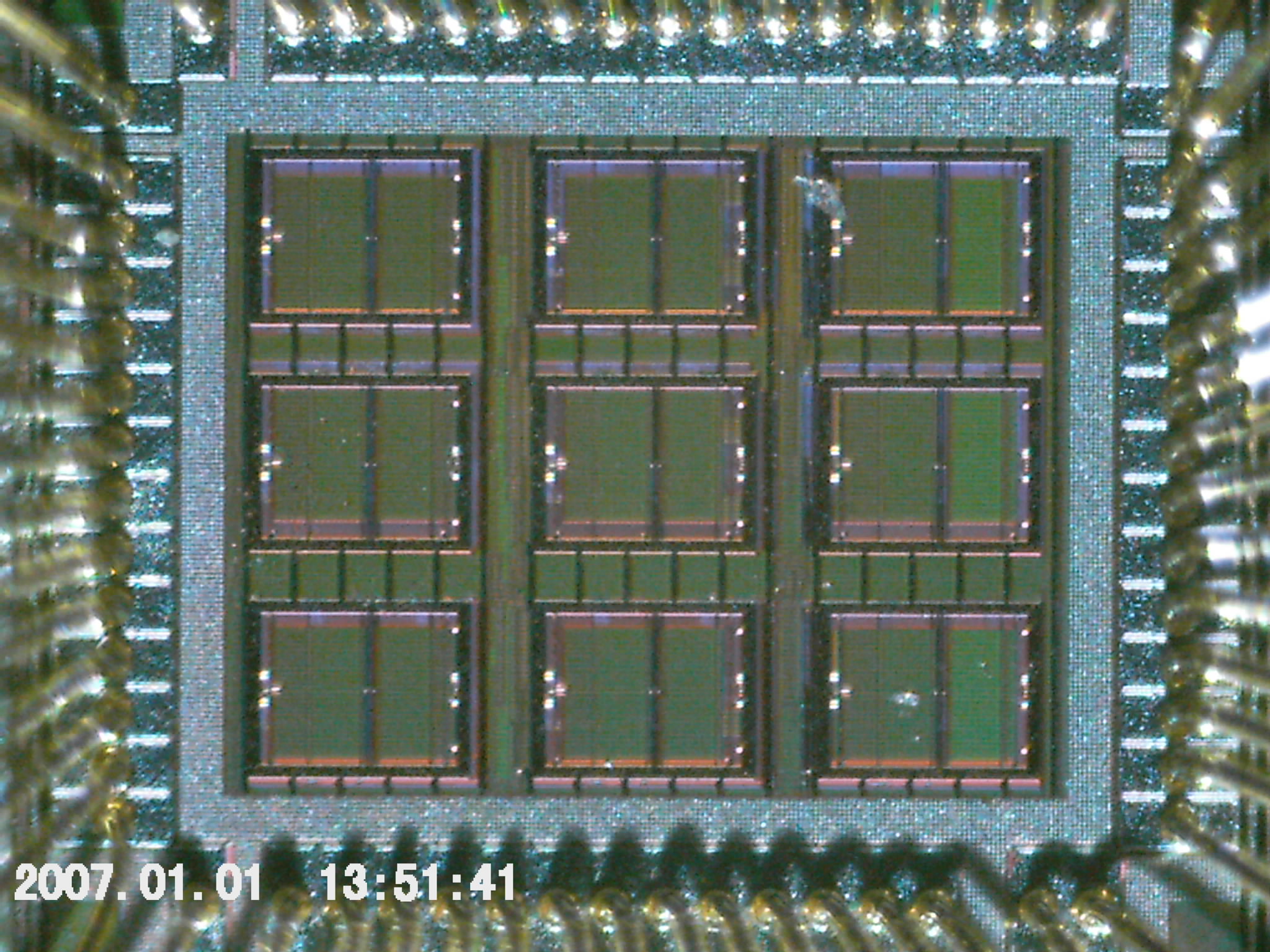}
\label{fig:protolayout3}}
}
\caption{Prototype: This chip has been fabricated in CMP run S65C8\_1 in Sept.2008, the date on the photo is a camera malfunction.}
\label{fig:proto}
\end{figure}

Figure~\ref{fig:protolayout3} shows the $3\times3$ prototype asynchronous FPGA which has been delivered by the foundry. This chip has been fabricated using ST Microelectronics
65 nm 7-layer process. The FPGA channel width is 8, and there are 9 I/Os for each side.

This FPGA has been laid out using a automatic flow as described in~\cite{chaudhuri-fpl07}.
The balanced place and route has been obtained through the following steps:
\begin{itemize}
\item
We first placed the switches, in a symmetric fashion as outlined in the layout schemes. 
\item
All channel segments are manually routed to achieve required balance.
\item
Configuration memory points and signals are placed/routed automatically,
because those resources are not sensitive.
\end{itemize}

Since size of the FPGA mainly depends on  size of switches and configuration memory points, and not limited by the routing area. Incorporation of
balanced routing does not result in an overhead in terms of area.

The prototype occupies an area of $1111.6 \mbox{ $\mu$m} \times 947.6 \mbox{ $\mu$m}$ in silicon and contains approximately 200,000 transistors.

\section{Experiments}
\label{sec:Experiments}

\begin{figure}
\centering
\includegraphics[width=0.48\textwidth]{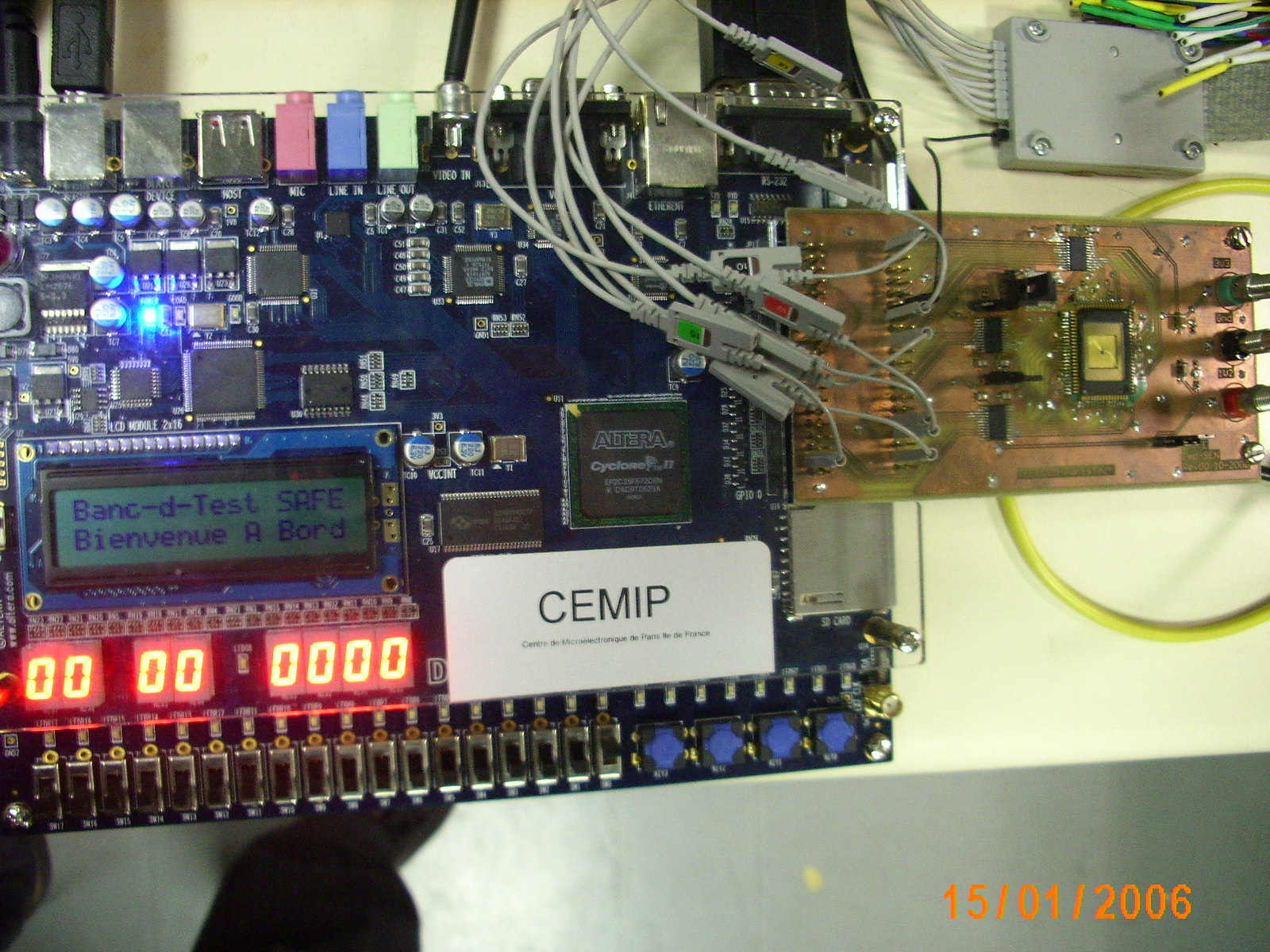}
\caption{The test setup with Altera DE2 board and a small PCB containing our Prototype.}
\label{fig:these_results_0}
\end{figure}
We have carried out two experiments. Figure~\ref{fig:these_results_0} shows the experimental setup. We used a DE2 board from Altera to
provide the test signals and acquisition of response from the prototype FPGA. The synchronous-to-asynchronous converters are implemented in
the DE2 board.
\subsection{Experiment 1: Configuration} 
\label{subsec:exp_config}


\begin{figure}
   \centering
   {
   \subfigure[The whole configuration trace of 4692 bits.]{
   \includegraphics[width=0.5\textwidth]{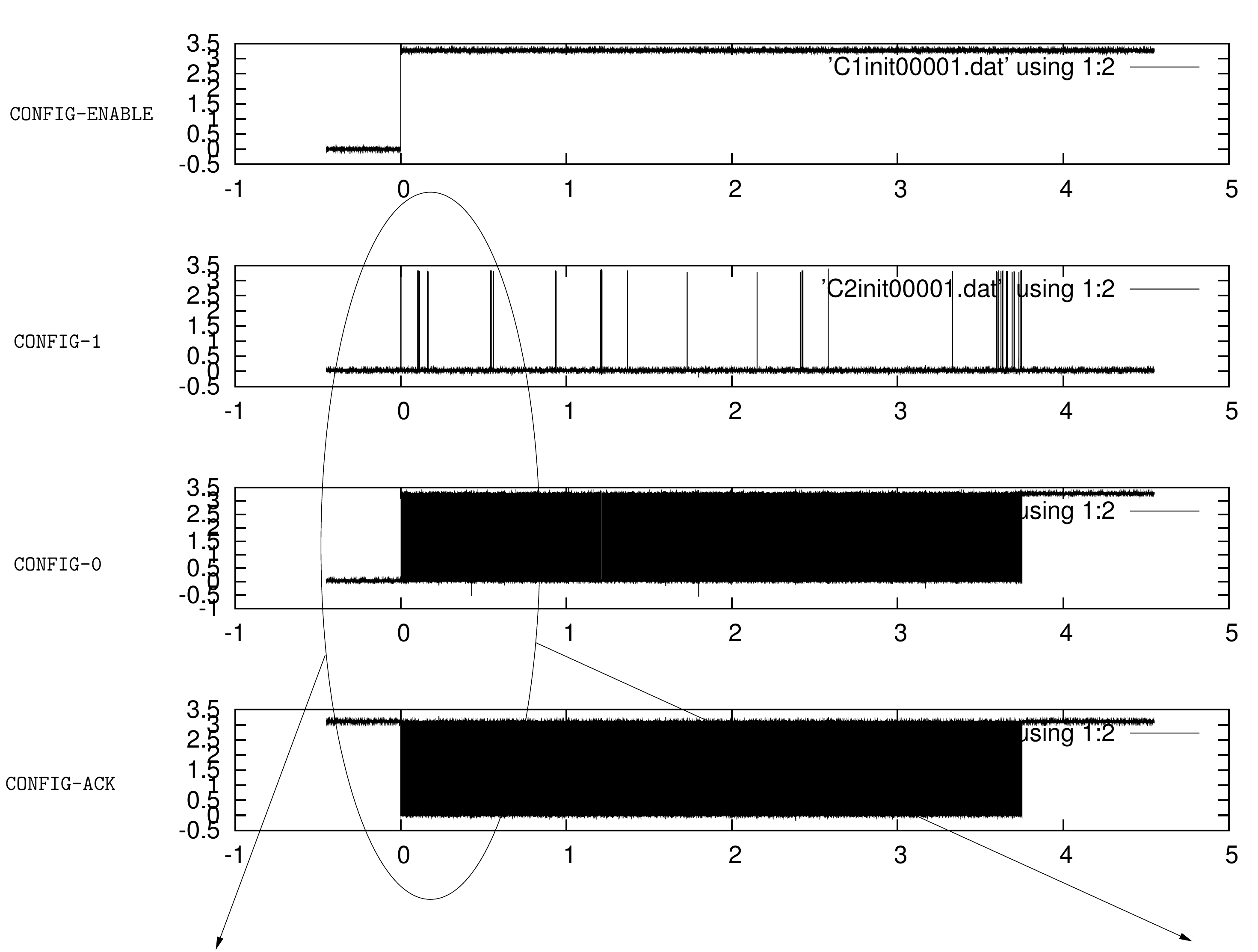}
   \label{fig:config_meas}}
   \subfigure[Zoomed view.]{
   \includegraphics[width=0.5\textwidth]{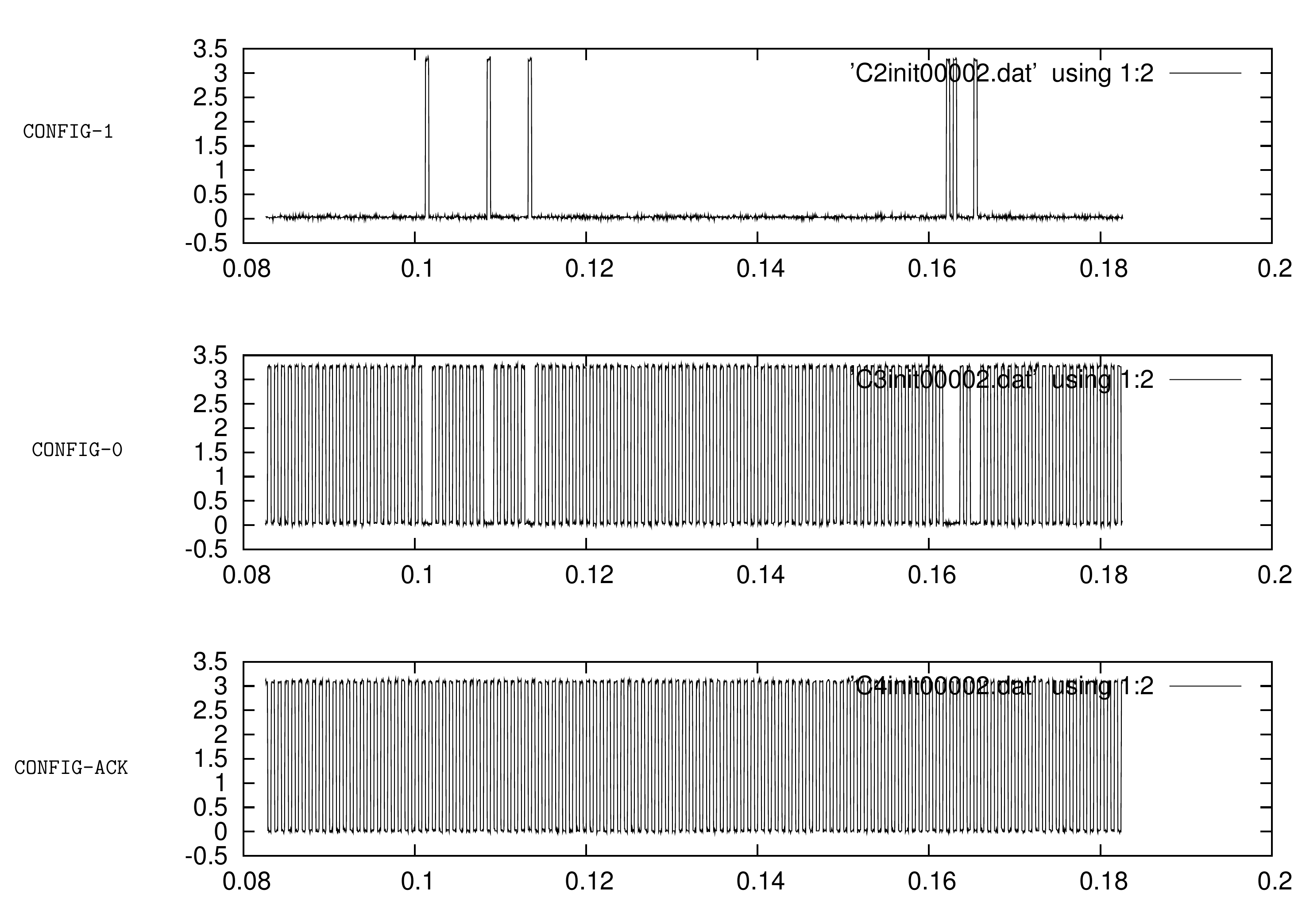}
   \label{fig:config_meas_zoom}}
}
   \caption{Measured trace of configuration signals.}
\end{figure}
The sequence for this experiment is the following:
\begin{itemize}
\item
First the asynchronous configuration chain inside SAFE is initialised, by putting the \emph{INIT} input of SAFE to `0' and then released after some time.
\item
The bitstream file generated by VPR~\cite{vpr} is loaded onto the driver board (an Altera DE2) RAM from a PC.
This bitstream is then converted to asynchronous $1$-out-of-$2$ coding
and sent to the \emph{configuration-0} and \emph{configuration-1} input signals of SAFE.
\item
The DE2 Board monitors the \emph{acknowledge} output from SAFE, and puts a new value in the configuration chain following 4-phase handshake.
It also counts the number of acknowledges received.
For a successful configuration it should receive exactly $4692$ acknowledgments.
\end{itemize}

We observe that when the bitstream contains only `0's the configuration is successful each time and with a very high speed.
We tested up to $50$~MHz, the DE2 board frequency.
When the bitstream contains `1's the configuration only succeeds at a low speed (around $10$~kHz).

Despite the fact that due to asynchronous coding `0's and `1's are equivalent in terms of transitions, we think that this behaviour might be due to a bug 
which we will be trying to explain with simulations in appendix~\ref{subsec:bug} at page~\pageref{subsec:bug}.
However, it does not hinder the test in real silicon of the routing strategy presented in Sec.~\ref{sec:Routing Architecture}.

%
%
%

\subsection{Experiment 2: Measuring Hop Mismatch}
\label{subsec:exp2}
\begin{figure}
   \centering
   {\subfigure[Measuring the difference between rail-0 and rail-1 for different lengths of rail-0.]{
   \includegraphics[width=0.45\textwidth]{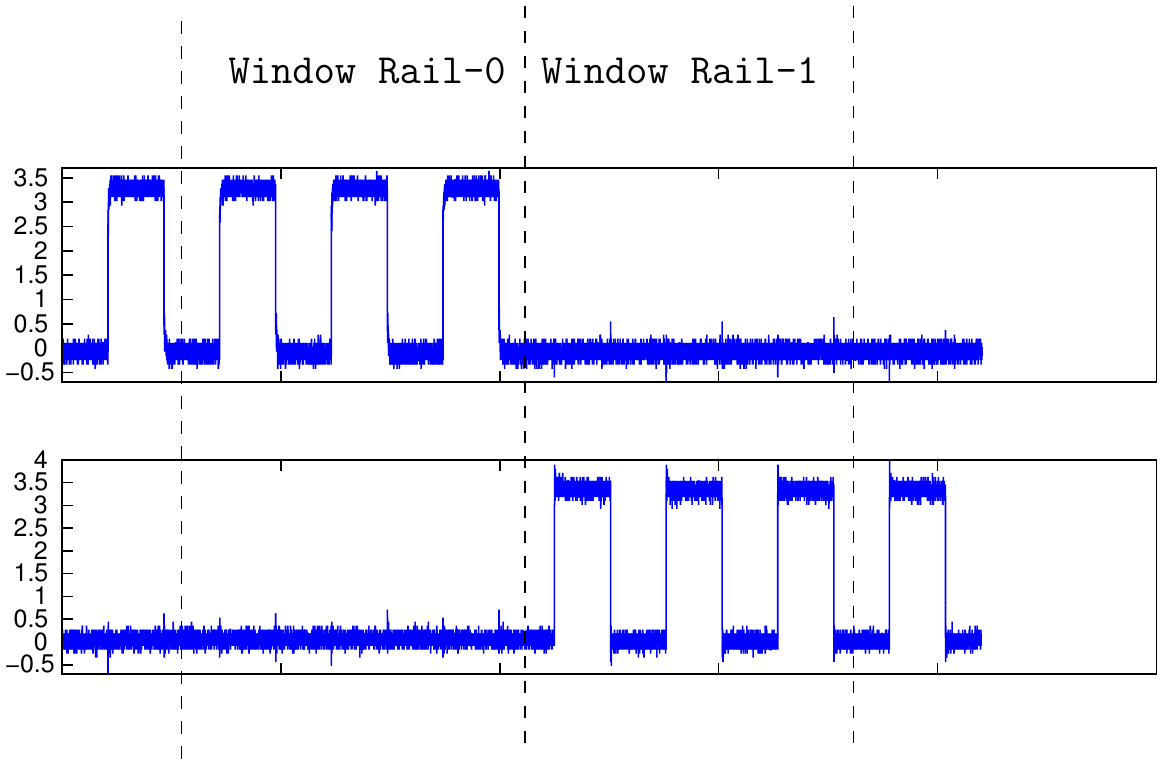}
   \label{fig:exp_hop_m}}
   \subfigure[Different pairs of dualrails to evaluate effect of hop-mismatch.]{
   \includegraphics[width=0.45\textwidth]{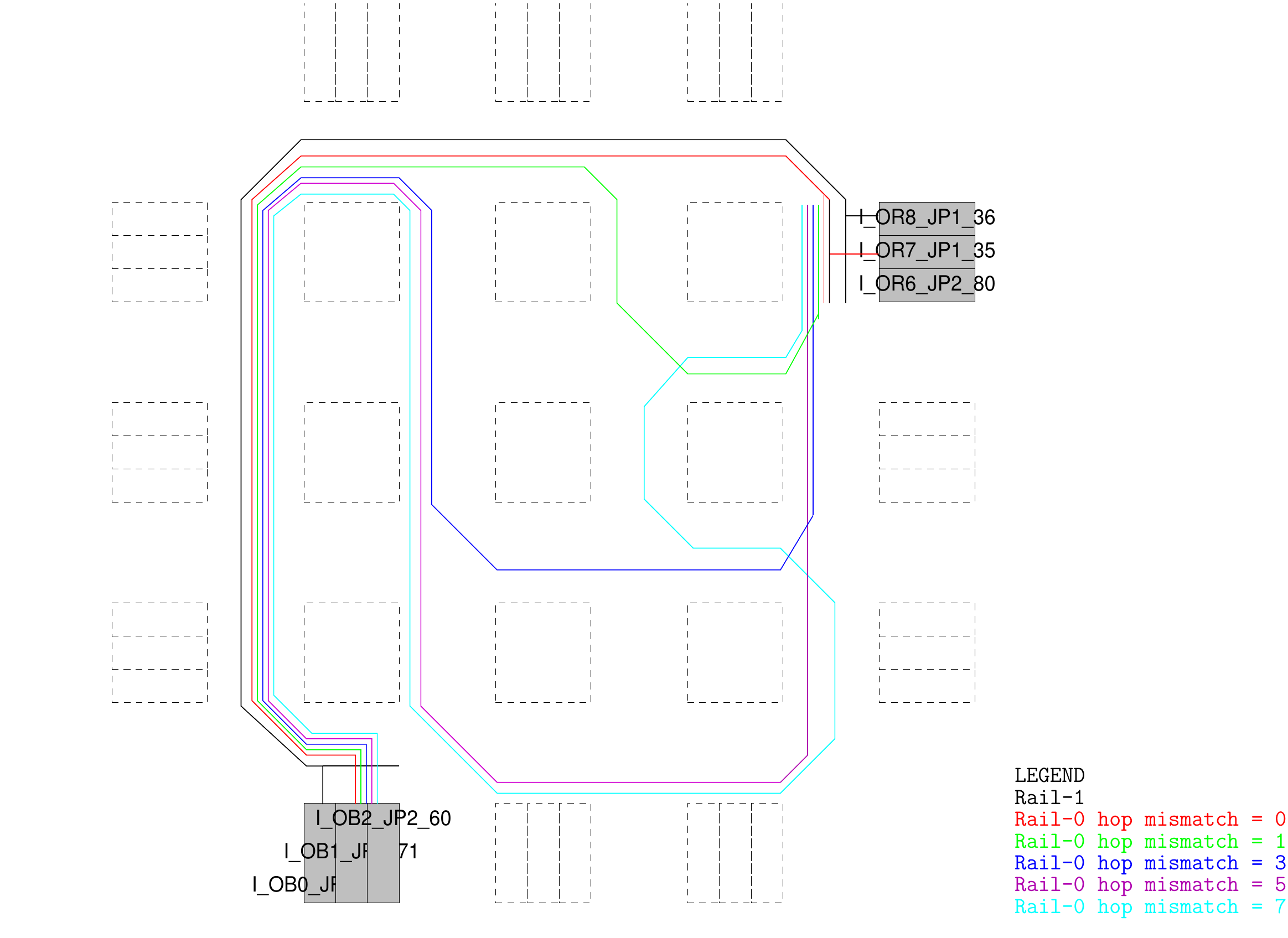}
   \label{fig:exp_hop_m1}}}
   \caption{Description of the experiment 2 ``Hop Mismatch''.}
\end{figure}
The goal of this experiment is to find out, the balancedness between the two rails constituting the dual-rail of asynchronous logic.
This is very important in terms of robustness against side-channel attack since any difference between these two rails will give out the data values to the attacker.
The power-constant methodology entirely depends on this balancedness.
Indeed, other biases, like the ``early propagation effect''~\cite{suzuki2006},
do not exist in a ``routing-only'' netlist.

Figures~\ref{fig:exp_hop_m} and \ref{fig:exp_hop_m1} describe the experimental setup. We route a dual-rail netlist in the FPGA. 
The route for RAIL-1 of this dual-rail stays constant over the experiment. The route for RAIL-0 is varied incorporating
(0,1,3,5,7) differences in hops w.r.t. RAIL-1, as shown in figure~\ref{fig:exp_hop_m1}. We send 4 pulses each for 
RAIL-1 and RAIL-0, and we measure the power consumption. This measurement is done with a separate trigger signal
encompassing $4+4$ pulses. Among the $4$ pulses we choose a window for the comparison. Let's say $W_1(t)$ is the trace for
RAIL-1 and $W_0(t)$ is the trace for RAIL-0 within this window.

The  balancedness between these two traces is then calculated as:
\begin{displaymath}
Balance = \frac{RMS(W_0(t))}{RMS(W_1(t))} \,,
\end{displaymath}
where RMS is the root mean square.
This ratio captures the difference of energy contained in each side-channel curve.

\begin{figure}
\centering
\includegraphics[width=0.48\textwidth]{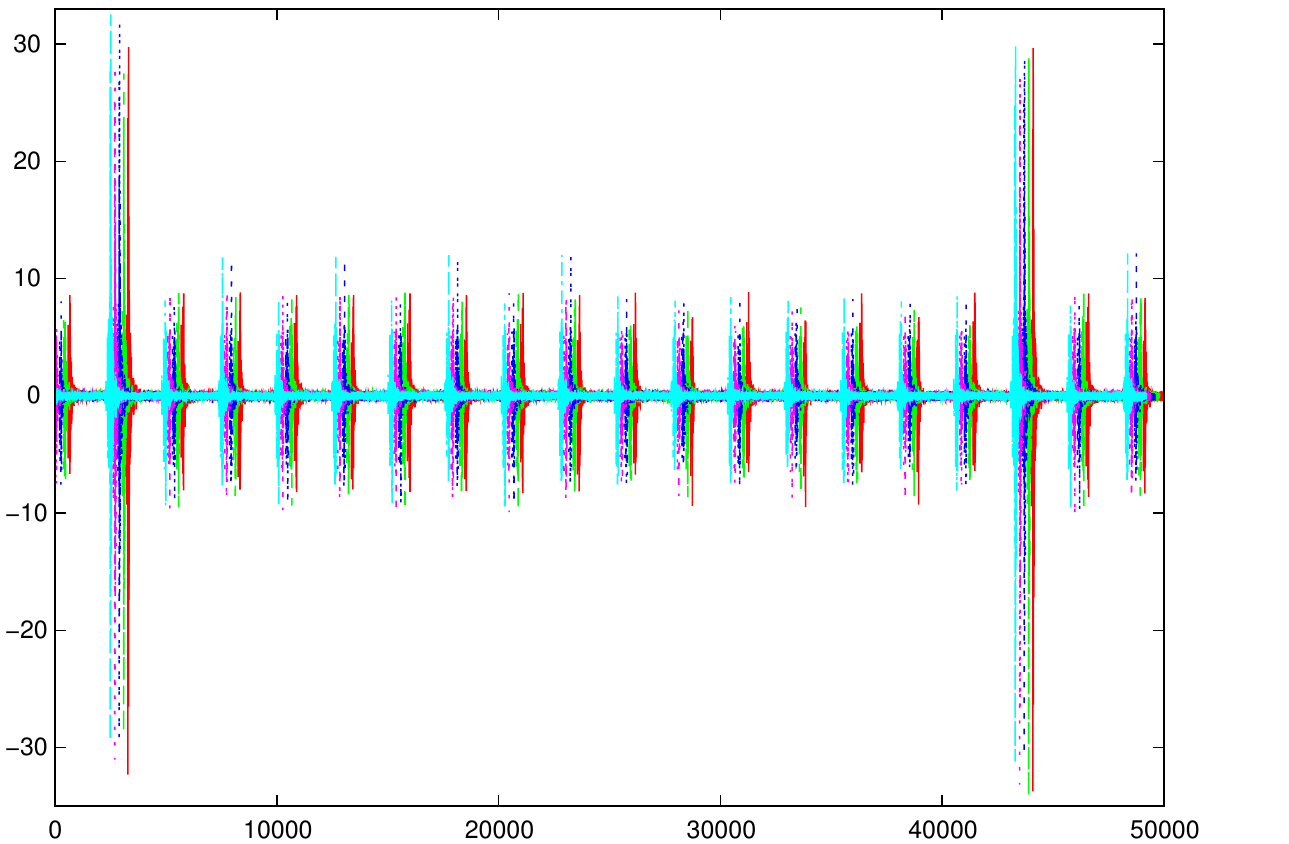}
\caption{The results for different values of Hop Mismatch (superimposed and staggered).}
\label{fig:these_results_1}
\end{figure}

The corresponding traces are shown in Fig.~\ref{fig:these_results_0}. We can see the triggering instants of the measurement.
The traces for various hop mismatch and the balancedness values are indicated in ~\ref{fig:these_results_1}. 
Figure~\ref{fig:these_results_0} shows the traces for different hop mismatch values superposed in a staggered fashion for comparison purpose.
They use the same color code as in fig.~\ref{fig:exp_hop_m1}.

\begin{center}
\setlength{\extrarowheight}{2pt}
\begin{tabular}{|c|c|}
\hline
Hop Mismatch & Balance \\
\hline\hline
0 & 1.106 \\
\hline
1 & 1.120 \\
\hline
3 & 1.158 \\
\hline
5 & 1.119 \\
\hline
7 & 1.173 \\
\hline
\end{tabular}
\end{center}


The results show that the hop mismatch definitely has an influence on the observed balancedness.
This deviation of the balancedness from one indicates that a bias exists;
such a bias is typically exploited by side-channel analyses, and therefore quantifies the implementation vulnerability.
One also notes that the unbalance is not strictly speaking linear with the hop mismatch.
\subsection{A note on Variability}
The results regarding "hop-mismatch" and "Balance of Power consumption" in subsection~\ref{subsec:exp2} can be interpreted as the superposition of 
a linear component and random components.
We can see that power consumption of a dual rail is largely proportional to the hop mismatch, so there is a strong linear
component of power consumption of dual-rails. However there is also a non-linear component as we can see from the result
for "hop mismatch =5". The authors suspect that this non-linear component is mainly due to variation. Notable causes of variation
in deep-submicron CMOS are~\cite{Takeuchi2009}
\begin{itemize}
\item
Systematic Process Variation
\item
Layout Dependent Variation
\item
Random Variation (Transistor Mismatch)~\cite{Pelgrom1989}
\end{itemize}
In this article we tried mainly to minimize the Layout dependent variation, and systematic process variation can be 
reduced by using matched transistor pairs. However the ever increasing random component of variation (with technology scaling)
could still be exploited by the attacker, and probably constitutes a fundamental limit to the anonymity from side-channel attackers.

\subsection{Secure Dual Rail Routing}

\begin{figure}[t]
\centering
\includegraphics[width=0.45\textwidth]{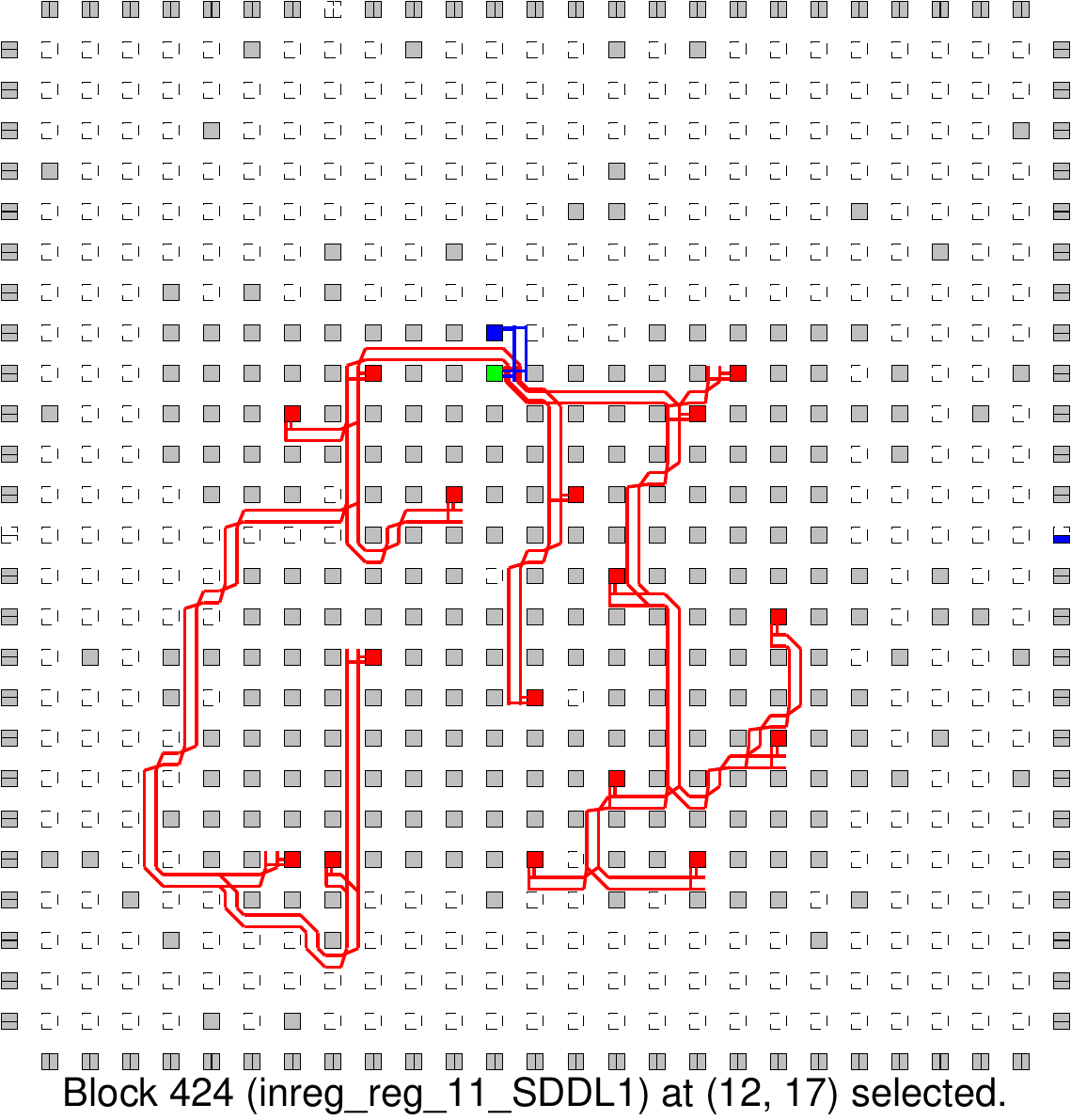}
\caption{Dual Rail routing Example}
\label{fig:routs_wddl}
\end{figure}
As we have seen the previous chapter, that there is a strong linear component in the effect of hop-mismatch, and power consumption balance
 between dual-rails, we should  guarantee balanced dual rail routing in CAD 
tools. We devised a simple dual rail routing algorithm where the routing tracks in FPGAs are divided into two domains, and each rail is then routed 
in separate domains. Because of the symmetrical domains, the routes for both rails are the same. However this will work only for homogeneous architectures,
such as tracks with unit lengths, and subset switchbox. In table~\ref{tab:dualroute} we show the results of dual rail routings for some netlists in the QUIP~\cite{quip}Benchmarks suite in a simple uniform architecture. The benchmarks are WDDL implementation of the netlists. As shown in table~\ref{tab:dualroute} we can 
guarantee a zero hop-mismatch routing with this technique. However this is only valid for simple homogeneous architectures, and needs considerable 
modification to be used in commercial architectures.
\begin{table*}
\begin{center}
\caption{Channel Width and Hop-Mismatch for Dual Rail Routing}
\label{tab:dualroute}
\begin{tabular}{|l|p{1cm}|p{1cm}|p{2cm}|p{1cm}|p{2cm}|}
\hline
\multicolumn{1}{|c}{Netlist} &
\multicolumn{1}{|c}{No. Nets} &
\multicolumn{2}{|c}{Breadth First} & 
\multicolumn{2}{|c|}{Dual Breadth First} \\ \hline
\multicolumn{1}{|c}{} &
\multicolumn{1}{|p{1cm}}{} &
\multicolumn{1}{|p{1.2cm}}{Channel Width} &
\multicolumn{1}{|p{2cm}}{Hop Mismatch /Dual-rail} &
\multicolumn{1}{|p{1.2cm}}{Channel Width} & 
\multicolumn{1}{|p{2cm}|}{Hop Mismatch /Dual-rail} \\ \hline

barrel16\_wddl	&626&15 &2.89 & 16 & 0 \\ \hline 	
barrel32\_wddl  &1482 &20 &2.78 &20 & 0 \\ \hline   
barrel64\_wddl  &3254 &23 &4.41 &22 & 0 \\ \hline
mux32\_16bit\_wddl&2964&12 &0.83 &12 &0 \\ \hline
mux64\_16bit\_wddl&5854&14 &0.51 &14 &0 \\ \hline
mux8\_128bit\_wddl&5932&11 &2.66 &12 &0 \\ \hline
mux8\_64bit\_wddl &2988&10 &2.25 &10 &0 \\ \hline
xbar\_16x16\_wddl &706&14 &1.59 &14 &0 \\ \hline

\end{tabular}
\end{center}
\end{table*}

\section{Conclusion}
\label{sec:Conclusion}
In this article we discussed the suitability of an asynchronous FPGA as a countermeasure to the physical cryptanalyses, and as
a prototyping device for such countermeasures. Intrinsic resistance of asynchronous circuits to faults injection, and power constant signalling
makes them good candidates for such countermeasures. Moreover because of their reconfigurable nature, it is possible to incorporate dynamic 
countermeasures along with static countermeasures. We believe a practical solution must use both of them to achieve highest level of security.
We presented approximate models of power consumption and delay on which our countermeasures are based, and defined our objectives for static 
countermeasures.

Keeping the prototyping role in mind, we presented a multi-style asynchronous PLB, and proposed a fine grain routing architecture, so that any 
$m$-out-of-$n$ coding and various asynchronous protocols can be mapped onto this architecture. 
As shown in various experiments throughout the article 
the power constant logical level protocols can not succeed without balanced interconnects. Layout statistics, and experiments on the extracted 
netlist from a prototype FPGA, presents the kind of balancedness in dynamic power consumption that can be achieved with the subset switchbox and 
the associated binary tree connection box. We also present a new physical implementation of the FPGA switch box called Tpair switchbox, which 
provides indiscernability in EM emission for the dual-rails routed through it.

Although the solutions proposed in actual layout assume bidirectional FPGA interconnect, we show how these solutions can be ported to other 
flavours of interconnect such as single-driver, both at the switchbox and connection box levels.
Finally we provide with some experimental results on $3\times 3$ prototype, although largely hindered by a bug in the circuit. However these test results 
will be of use to future designs. We carry out a profiling of power consumption for different values of hop mismatch and we see a clear dependence.
The experiment on extracted netlist shows the very high speed of the asynchronous configuration chain $\sim$1.6~GHz, and we also verified the functionality 
of this in silicon at a lower speed, again because of the bug which will be corrected in the future designs.

In this paper  we mainly concentrated on layout dependent (geometric) variations in dual-rails, which is only one component of the variations
in power consumption between two rails. However from experimental results we discover significant other components in power consumption balance.
Hence as future research directions, we would like to propose nullifying the effect of CMOS variation (by taking alternate routes in a random fashion)
and from other sources of variation from rail to rail.
We would also like to stress the importance of CAD algorithms  in physically securing the application, such as, 
automatically routing dual-rails through the FPGA in a balanced way.
These important issues are the main future research challenges.

\bibliographystyle{plain}
\bibliography{acmtrets,dpa_fpga_state_of_the_art,fpl_refs,sca}

\appendix
\section{Possible Bug in the Fast Asynchronous Configuration Chain}
\label{subsec:bug}
Due to the problems encountered during the configuration as explained in subsection~\ref{subsec:exp_config}, we investigated into the cause of this anomaly.
We did the simulation of a single 6-input LUT inside the PLB. The implementation of this LUT is explained in subsection~\ref{sec:Balanced LUT Implementation}.
In figure~\ref{fig:config_bug} we provide a more detailed view of the LUT configuration chain and switches. In actual silicon the memory points are connected 
to the LUT output through Transmission Gates (denoted as pass transistors in fig.~\ref{fig:config_bug}). The LUT inputs are decoded in a such a way that only 
one among these parallely connected Transmission Gates is ``ON'', depending on the input value, and the corresponding memory point should appear at the output.

However when the inputs go through a transition, there is a temporary short-circuit between the two concerned Transmission Gates (TGs). Because of the bidirectional 
nature of TGs, this disturbs the configuration chain itself. One LUT memory point is written into another one.

We did the same simulation with tri-state buffers instead of Transmission Gates, in this case the output changes according to the inputs and memory points.
We think that this could be a possible cause of the anomaly during the configuration of the FPGA ``SAFE''. In actual silicon, during the configuration, the LUT inputs 
are not forced to '0' or '1'. If the inputs go through parasitic transitions during configuration, this will introduce invalid state $('1','1')$ in the configuration chain leading to random behaviour.

Further investigations, and testing is going on at TIMA, Grenoble, so that the FPGA can be used for basic testing, and this simulation results will be taken into 
account during the next tape out.


%

\end{document}